\documentclass[fleqn,usenatbib]{mnras}

\usepackage{newtxtext,newtxmath}

\usepackage[T1]{fontenc}

\usepackage{enumitem}

\DeclareRobustCommand{\VAN}[3]{#2}
\let\VANthebibliography\thebibliography
\def\thebibliography{\DeclareRobustCommand{\VAN}[3]{##3}\VANthebibliography}

\usepackage{graphicx}	
\usepackage{amsmath}	
\usepackage{multirow}
\usepackage{pdflscape}
\usepackage[flushleft]{threeparttable}
\usepackage{caption}
\newcommand{\angstrom}{\text{\AA}}

\newcommand{\cii}{[C\,{\sc ii}]}

\newcommand{\Tdust}{T_{\rm d}}

\newcommand{\oiiil}{[O\,{\sc iii}] 88\,$\mu{\rm m}$}
\newcommand{\ciil}{[C\,{\sc ii}] 158\,$\mu{\rm m}$}

\title[Hot n Cold]{Realistic Multi-temperature Dust: How Well Can We Constrain the Dust Properties of High-redshift Galaxies?}

\author[Sommovigo \& Algera]{Laura Sommovigo$^{1}$\thanks{E-mail: lsommovigo@flatironinstitute.org} and 
Hiddo Algera$^{2}$\thanks{E-mail: hsbalgera@asiaa.sinica.edu.tw} \\ 
$^{1}$ Center for Computational Astrophysics, Flatiron Institute, 162 5th Avenue, New York, NY 10010, USA \\
$^{2}$ Institute of Astronomy and Astrophysics, Academia Sinica, 11F of Astronomy-Mathematics Building, No.1, Sec. 4, Roosevelt Rd, Taipei 106319, Taiwan, R.O.C.}

\date{Accepted XXX. Received YYY; in original form ZZZ}

\pubyear{2025}
\begin{document}
\label{firstpage}
\pagerange{\pageref{firstpage}--\pageref{lastpage}}
\maketitle

\begin{abstract}
Determining the dust properties of high-redshift galaxies from their far-infrared continuum emission is challenging due to limited multi-frequency data. As a result, the dust spectral energy distribution (SED) is often modeled as a single-temperature modified blackbody.
We assess the accuracy of the single-temperature approximation by constructing realistic dust SEDs using a physically motivated prescription where the dust temperature probability distribution function (PDF) is described by a skewed normal distribution. This approach captures the complexity of the mass-weighted and luminosity-weighted temperature PDFs of simulated galaxies and quasars, and yields far-infrared SEDs that match high-redshift observations. We explore how varying the mean temperature ($\bar{T}_\mathrm{d}$), width, and skewness of the temperature PDF affects the recovery of the dust mass, IR luminosity, and dust emissivity index ($\beta_\mathrm{d}$) at $z=7$.
Fitting the dust SEDs with a single-temperature approximation, we find that dust masses are generally well-recovered, although they may be underestimated by up to $0.6\,\mathrm{dex}$ for broad temperature distributions with a low $\bar{T}_\mathrm{d}\lesssim40\,\mathrm{K}$, as seen in some high-redshift quasars and/or evolved galaxies. IR luminosities are generally recovered within the $1\sigma$ uncertainty ($\lesssim0.3\,$dex), except at $\bar{T}_\mathrm{d}\gtrsim80\,$K, where the peak shifts well beyond ALMA's wavelength coverage.
The inferred dust emissivity index is consistently shallower than the input one ($\beta_\mathrm{d}=2$) due to the effect of multi-temperature dust, suggesting that a steep $\beta_\mathrm{d}$ may probe dust composition and grain size variations. With larger galaxy samples and well-sampled dust SEDs, systematic errors from multi-temperature dust may dominate over fitting uncertainties and should thus be considered.
\end{abstract}




\section{Introduction}
\label{sec:introduction}
In the last decade, observations with the Atacama Large Millimeter/submillimeter Array (ALMA) have demonstrated that dust is present -- and perhaps even commonplace -- in the epoch of reionization (e.g., \citealt{watson2015,bowler2018,bowler2024,marrone2018,hashimoto2019,tamura2019,harikane2020,bakx2021,bakx2024,fudamoto2021,bouwens2022,inami2022,schouws2022,witstok2022,witstok2023_beta,algera2023,hashimoto2023_rioja,tripodi2023,tripodi2024,vanleeuwen2024}). To understand how all this dust came to be, and how much star formation it obscures, characterizing these distant dust reservoirs is paramount. This requires accurate measurements of galaxy dust masses and temperatures, which in turn necessitate multi-band sampling of the dust spectral energy distribution (SED). 

From an observational perspective, the shape of the dust SED of high-redshift galaxies is often approximated by that of a single-temperature optically thin modified blackbody (MBB; e.g., \citealt{bakx2021,witstok2022,algera2024}), which depends on three parameters: the dust mass $M_{\rm d}$, an effective dust temperature $T_{\rm d,MBB}$, and the dust emissivity index $\beta_\mathrm{d}$. In the aforementioned approximation, the measured continuum flux density $F_{\rm \nu}$ observed against the CMB at frequency $\nu$ can be written as \citep[see e.g.,][]{dacunha2013,Kohandel19}
 \begin{equation}\label{flux_eq}
F_{\rm \nu} =  g(z) M_{\rm d} \kappa_{\nu} [B_{\nu}(T_{\rm d,MBB})-B_{\nu}(T_{\rm CMB})],
\end{equation}
where $g(z) = {(1+z)}/{d_L^2}$, $d_{\rm L}$ is the luminosity distance to redshift $z$, $\kappa_{\rm \nu}\propto \nu ^{\beta_d}$ is the dust opacity, $B_\nu$ is the black-body spectrum, and $T_{\rm CMB}(z)$ is the CMB temperature\footnote{$T_{\rm CMB}(z)=T_{\rm CMB,0}(1+z)$, with $T_{\rm CMB,0}=2.7255\, \mathrm{K}$ \citep{Fixsen09}} at redshift $z$.

At $z\gtrsim5$, most attention has been directed to measuring $M_{\rm d}$ and $T_{\rm d,MBB}$ (e.g., \citealt{hashimoto2019,faisst2020,witstok2022,mitsuhashi2024}), while it is common to fix the value of $\beta_\mathrm{d} \approx 1.5 - 2$ based on observations of galaxies at lower redshift (e.g., \citealt{dacunha2021,cooper2022,bendo2023,liao2024}).
Even if the effective dust temperature inferred from SED fitting per se might not be a key quantity for galaxy formation theory -- nor a physically straightforwardly interpretable one -- it majorly affects derived galaxy properties such as the dust content and obscured SFR $\propto L_\mathrm{IR} \propto M_\mathrm{d} T_{\rm d,MBB}^{4+\beta_\mathrm{d}}$ \citep[see e.g.,][for extensive discussions on the topic]{sommovigo2020,sommovigo2022b}.
For instance, due to the assumption of a Milky Way-like cold dust effective temperature $T_{\rm d,MBB}=25\ \mathrm{K}$ out to high redshift, the dust masses inferred for galaxies with a single continuum detection such as the targets of large programs like ALPINE \citep{lefevre2020,faisst2020,bethermin2020} were initially overestimated \citep{2021A&A...653A..84P}, requiring dust yields mostly incompatible with known dust production channels -- mainly AGB stars, supernovae (SNe), and growth in the interstellar medium \citep[see the discussions in][]{sommovigo2022b,Choban2024b}. These assumptions also imply low obscured SFR fractions at high redshift, which seemed to be consistent with the low visual attenuation and blue UV slopes displayed by the (UV-selected and thus inherently biased) high-$z$ galaxies. However, recent studies (both theoretical and observational) have shown that due to warmer temperatures $T_\mathrm{d,MBB}\sim 40-45\ \mathrm{K}$ obscured SFR fractions exceeding $\sim 80\%$ can be found even in UV bright systems out to $z\sim8$ \citep{bakx2021,inami2022,algera2023,mitsuhashi2024,Valentino24}.

Several recent studies have attempted to constrain the FIR SEDs of high-redshift galaxies with multi-band ALMA follow-up (e.g., \citealt{bakx2020,bakx2021,harikane2020,sugahara2021,witstok2022,akins2022,algera2024,algera2024b,mitsuhashi2024}). These works have highlighted a large spread in measured effective dust temperatures at $z\approx6-8$, ranging from $T_{\rm d,MBB}\approx30\,$K \citep{witstok2022,algera2024} to $T_{\rm d,MBB}\gtrsim85\,$K (\citealt{tamura2019,bakx2020,jones2024}). While some of this observed spread is likely due to the limited sampling of the dust SEDs at just two or three distinct wavelengths (c.f., the discussion in \citealt{bakx2021,algera2024b}), it also reflects intrinsic variation in the underlying galaxies themselves \citep[such as metallicity, depletion time, sSFR and gas column density; see e.g.,][]{Hirashita22,sommovigo2022,sommovigo2022b}. 

Despite the growing number of multi-band -- and specifically high-frequency -- observations constraining the value of the effective dust temperature in high-redshift galaxies, it is a well-known fact both theoretically and observationally that physical dust temperatures are not just a single value. Observations of local galaxies, for instance, highlight significant spatial variations in their dust temperatures (e.g., \citealt{helou1986,utomo2019,chiang2023}), casting doubt on the validity of single-$T_{\rm d,MBB}$ models.
Spatial variations of the dust temperature have been investigated out to high redshift in a handful of bright and/or lensed sources. \cite{akins2022} constrain the radially-averaged dust temperature profile for the lensed galaxy A1689-zD1 at $z=7.13$, finding $\pm 10\ \mathrm{K}$ variation in $T_{\rm d,MBB}$ from central SF regions (warmer) to galactic outskirts (colder). Even larger variations are found by \cite{Tsukui_2023} in one of the brightest sub-millimeter sources known at $z>4$, a quasar host galaxy at $z=4.4$. They find that the best fir FIR SED is composed of a warm, AGN-heated component at $T_{\rm d,MBB} = 87\ \mathrm{K}$ and a colder one at $T_{\rm d,MBB} = 57\ \mathrm{K}$ (see also \citealt{fernandez-aranda2025,meyer2025} for similar recent studies). 
Now that multi-band ALMA coverage is becoming more widespread -- and that shorter wavelength IR instruments such as \textit{PRIMA} \citep{moullet2023} are being proposed and actively discussed -- it is increasingly important to consider potential higher-order effects related to multi-temperature dust components. 

On the theoretical side, distributions of the \textit{physical} dust temperature $T_{\rm d}$ are readily available and do not need to be inferred from SED fitting procedures. Analytical models \citep[e.g.,][]{sommovigo2020,sommovigo2022b,Hirashita22} and radiative-transfer post-processed cosmological zoom-in simulations \citep[e.g.,][]{behrens2018,liang2019,pallottini2022,lower2024} find stark spatial variations  ($\sim 100\,\mathrm{K}$) in the dust temperatures of galaxies in the early Universe (for quasar hosts see also \citealt{diMascia22}). The consensus is that in early galaxies, highly star-forming regions are highly turbulent and pressurized, thus denser and more heavily obscured, and longer-lived against feedback effects (Giant Molecular Cloud, GMC, lifetimes are predicted to be as large as $10$ Myr, vs. typical $2-3$ Myr locally; \citealt{sommovigo2020,Menon_2024,somerville2025}). Within these dense GMCs, dust is expected to attain $\geq 2 \times$ warmer temperatures than in typical local clouds ($T_{\rm d}\sim 60-100$ K, see e.g., \citealt{behrens2018,liang2019,sommovigo2020,pallottini2022}). 
In the local Universe, warm/hot dust pockets are also sparsely observed in the inner parts of some massive stars' HII regions \citep{2007ApJ...660..346P}; however, most of the dust budget is generally located in the outer photodissociation region (PDR), attaining temperatures closer to the average quoted Milky Way average value ($\sim 25$ K, see \citealt{Anderson12}). On top of dust in GMCs generally being hotter, the CMB also sets a warmer temperature floor at high redshift (e.g., $T_{\rm CMB} = 21.8$ K at $z = 7$) affecting the overall dust temperature distribution (albeit to only a modest degree). 

In brief, we know that the effective dust temperature $T_{\rm d, MBB}$ inferred for (mostly) unresolved galaxies is a non-trivial combination of a likely colder mass-weighted average temperature characterizing the diffuse ISM, and warmer luminosity-weighted average temperature probing hot, dense star-forming regions.  
How then does the observationally inferred temperature $T_{\rm d, MBB}$ compare to the underlying temperature distribution for varying shapes of said distribution? And to what extent does the assumption of a single-$T_{\rm d, MBB}$ fit systematically bias the inferred dust masses and IR luminosities -- and thus dust-obscured SFRs -- of distant galaxies? 

In this work, we aim to address these questions by constructing mock galaxy dust SEDs starting from a realistic distribution of underlying physical dust temperatures, and fitting these with single-temperature models as would be applied to actual (sub-)millimeter observations. We describe our method of constructing and fitting these mock SEDs in Section \ref{sec:methods} and present our results in Section \ref{sec:results}. We discuss these findings in Section \ref{sec:discussion} and present our conclusions in Section \ref{sec:conclusions}. Throughout this work, we adopt a standard $\Lambda$CDM cosmology, with $H_0=70\,\text{km\,s}^{-1}\text{\,Mpc}^{-1}$, $\Omega_m=0.30$ and $\Omega_\Lambda=0.70$. 

\begin{figure*}
    \centering
    \includegraphics[width=0.995\textwidth]{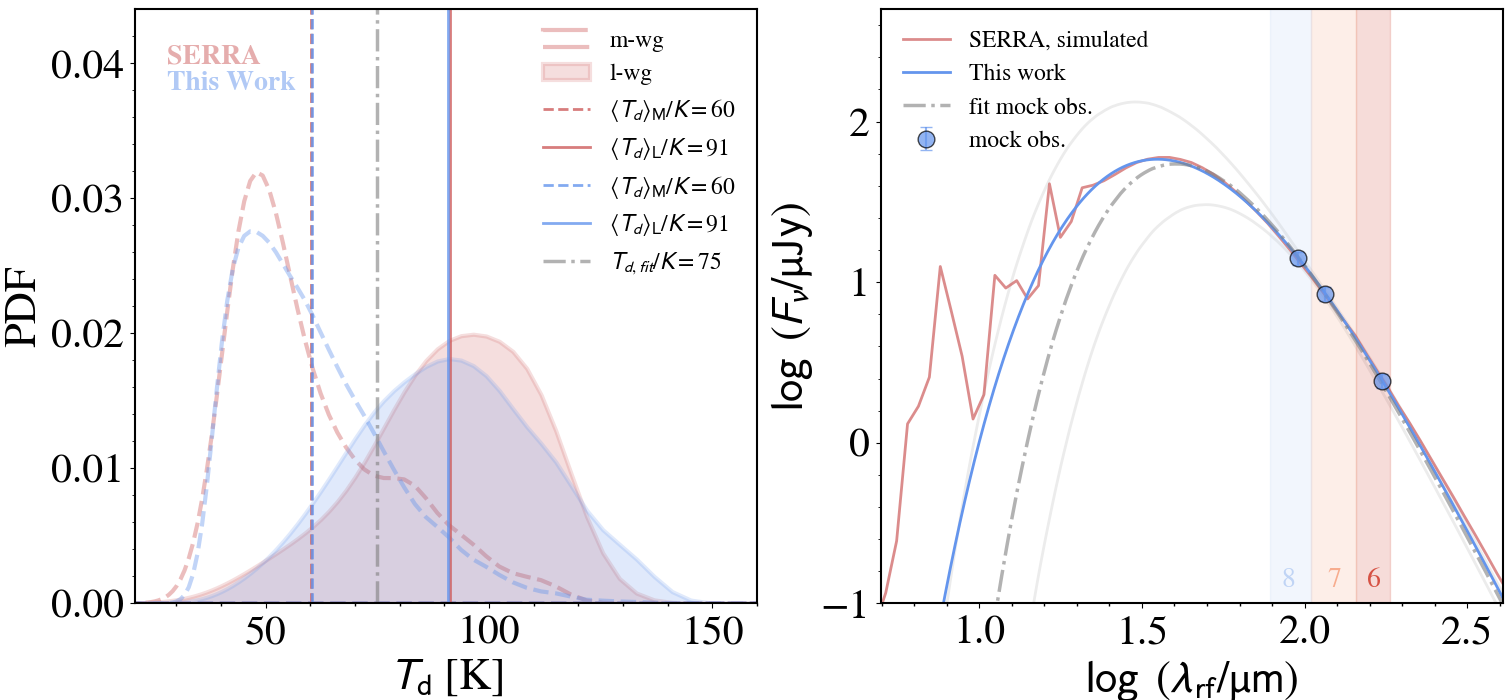}
    \caption{\textbf{Our simple analytical model can reproduce the mass- and luminosity-weighted dust temperature PDFs and IR SED of a simulated high-redshift galaxy.} \textit{Left panel}: The PDF of the dust temperature in the SERRA simulated $z=6.67$ galaxy Zinnia described in \citet{sommovigo2021,pallottini2022}. The pink dashed (solid) line shows the mass-weighted (luminosity-weighted) PDF and mass-weighted (luminosity-weighted) temperature $\langle T_{\rm d} \rangle_\mathrm{M}$ ($\langle T_{\rm d} \rangle_\mathrm{L}$) whereas the grey dot-dashed line shows the best-fit solution from the single-temperature modified blackbody fit (see right panel). The blue lines show the mass-weighted (dashed) and luminosity-weighted (solid) PDF and temperatures obtained with our analytical model assuming a skewed normal distribution for $\Tdust$ with $\bar{T}_{\rm d}=60\ \mathrm{K}$, $\sigma_T=16\ \mathrm{K}$, $\zeta=0.98$. The dust mass is set to the total dust mass of the simulated galaxy $M_{\rm d, tot}/M_{\odot} = 10^{5.33}$. \textit{Right Panel}: The IR SEDs corresponding to the PDFs shown in the left panel (in pink the simulated galaxy, in blue our model). The grey dash-dotted line (semi-transparent grey solid lines) represents the best-fitting (16th and 84th percentile) single-temperature modified blackbody curve for the mock observations in ALMA bands 6, 7, and 8 (blue points) extracted from our modeled SED. The MCMC fitting procedure is described in Sec.~\ref{sec:methods_fit_mockALMA}.}
    \label{fig:PDFandSED_SERRAlike}
\end{figure*}

\section{Methods}\label{sec:methods}

The goal of this work is to create a simple analytical model capable of parameterizing the multi-temperature dust distributions expected in galaxies in the Epoch of Reionization. We then use this model to create realistic dust SEDs, and subsequently fit them with single-temperature models as an observer would do. The details of this process are given below.

\subsection{A simple model for multi-temperature dust distributions }
In constructing the dust temperature distributions of distant galaxies, we draw inspiration from the SERRA cosmological zoom-in simulations \citep{pallottini2022}. 
We assume that the dust temperature distribution within galaxies follows a skewed normal distribution. Such an assumption is grounded in the central limit theorem and the asymmetry introduced in the $T_{\rm d}$ distribution by two key factors: the i) temperature floor set by the CMB, and ii) the hot dust temperatures ($T_{\rm d}\approx 100\ \mathrm{K}$) associated with actively star-forming regions \citep[][]{behrens2018,liang2019,sommovigo2020}. Such warm/hot dust pockets are expected to be widespread in high-z, young and bursty galaxies \citep{behrens2018,sommovigo2020,pallottini2022}, as also suggested by recent observations \citep{bakx2020,akins2022}. Furthermore, the hard upper limit on the dust temperature is dictated by the sublimation temperature of silicate and carbonaceous grains, which reaches up to $1200$ K and $2100$ K, respectively \citep{Kobayashi_09}.

\begin{figure*}
    \centering
    \includegraphics[width=0.999\textwidth]{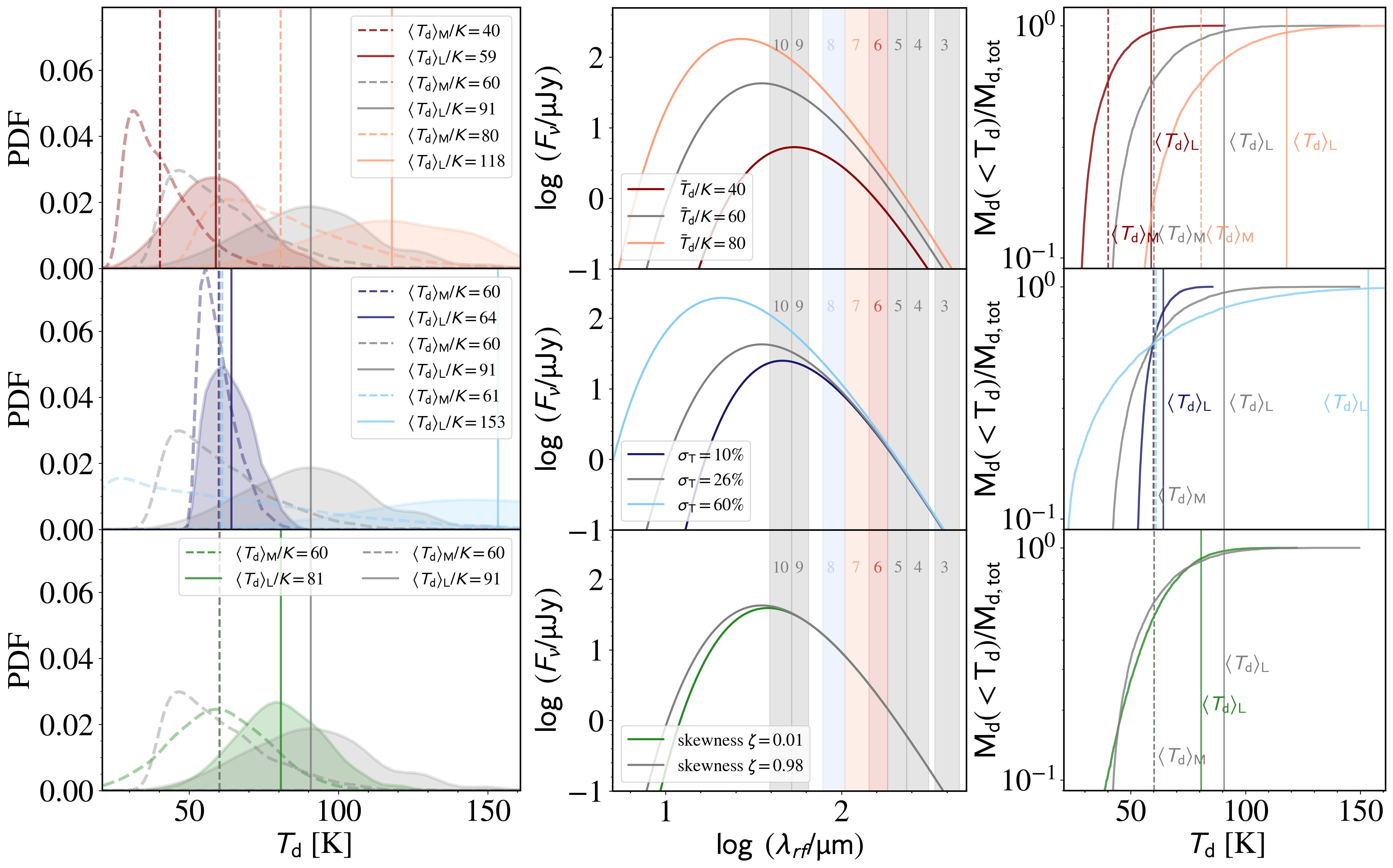}
    \caption{\textbf{The effect of varying our three model parameters on the dust temperature PDFs and IR SEDs of high-redshift galaxies.} \textit{Left and Central panels}: Dust temperature PDF and corresponding SED for different choices of the free parameters ($\bar{T}_{\rm d}$; $\sigma_T$; $\zeta$) in our model.     In the top row, we vary the mass-weighted temperature $\bar{T}_{\rm d}=(40,60,80)\ \mathrm{K}$, in the central row we change the variance $\sigma_T/\bar{T}_{\rm d}=(0.1,0.26,0.6)$, and in the lower one the skewness $\zeta=(0.01,0.98)$.
    The "fiducial" case reproducing Zinnia (shown in Fig. \ref{fig:PDFandSED_SERRAlike}) is shown in grey across all panels. For the dust temperature PDFs (leftmost panels) we use the same representation as in Fig. \ref{fig:PDFandSED_SERRAlike}, with the mass-weighted $T_{\rm d}$ PDFs represented by dashed lines, and the luminosity-weighted ones by solid, filled contours. For the SEDs (central panels), in addition to the ALMA bands (6,7,8) used in the "basic ALMA" configuration, we also show all the ALMA bands included in the "super ALMA" configuration (Section \ref{sec:methods_construct_mockALMA}).
 \textit{Right panels}: Fraction of the dust total mass located below a given dust temperature $M_{\rm d}(<T_{\rm d})/M_{\rm d,tot}$ for the choices of free parameters shown in the central and leftmost panels. The vertical dashed (solid) lines mark the mass-weighted (luminosity-weighted) $T_{\rm d}$ in each case.} 
    \label{fig:Expl_Param}
\end{figure*}

Under the skewed-normal assumption, the dust temperature Probability Distribution Function (PDF) is expressed as:
\begin{equation}\label{TdPDF}
    f(\Tdust) = \frac{2}{\omega} \phi \left( \frac{\Tdust - \xi}{\omega} \right) \Phi \left( \alpha \frac{\Tdust -\xi}{\omega} \right)
\end{equation}
where $\phi$ is the standard normal distribution and $\Phi(x) = 1/2\ [ 1 + erf(x / \sqrt{2}) ]$. We account for the effect of CMB heating by applying the following correction to the dust temperature PDF \citep{dacunha2013}:
\begin{equation}
    \label{eq:Tcmb}
    T_{\rm d}' = \{T_{\rm d}^{4+\beta_{\rm d}}+T_{\rm CMB,0}^{4+\beta_{\rm d}}[(1+z)^{4+\beta_{\rm d}}-1]\}^{1/(4+\beta_{\rm d})},
\end{equation}
where $T_{\rm CMB,0}=2.725\ \mathrm{K}$ is the CMB temperature at redshift $z=0$ and $\beta_{\rm d}=2$ is the adopted dust opacity index, valid for Milky-Way like dust \citep{weingartner2001}. 

The PDF in eq. \ref{TdPDF} has three free parameters: the location ($\xi$), scale ($\omega$), and shape ($\alpha$). These somewhat abstract parameters can be translated into physically interpretable quantities by expressing them as a function of the mean dust temperature ($\bar{T}_{\rm d}$), standard deviation ($\sigma_T$), and skewness ($\zeta$) of the dust temperature PDF as:
\begin{align}
    \xi (\bar{T}_{\rm d}, \sigma_T, \zeta) & = \bar{T}_{\rm d} - \omega \  \delta\  \sqrt{2/ \pi}\\
    \omega (\sigma_T, \zeta) & = \sigma_T / \sqrt{1-2\ \delta^2/\pi}\\
    \alpha (\zeta) & = \delta / \sqrt{1-\delta^2} 
\end{align}
where
\begin{equation}
    \delta(\zeta) = \sqrt{ \frac{\pi}{2}\ \frac{\zeta^{2/3}} { \zeta^{2/3} + ( \frac{4-\pi}{2} )^{2/3} } } 
\end{equation}
These parameters ($\bar{T}_{\rm d}$, $\sigma_T$, and $\zeta$) can be linked to global physical galaxy properties. For instance, since we assign the mass in each $T_{\rm d}$ bin according to the PDF in eq.~\ref{TdPDF}, the mean temperature $\bar{T}_{\rm d}$  corresponds to the mass-weighted dust temperature. In the following, we use $\bar{T}_{\rm d}$ when discussing the mass-weighted dust temperature in the context of an input to our model while using $\langle T_{\rm d} \rangle_{\rm M}$ when discussing it as a physical temperature. We distinguish between the two as we also compare with simulation outputs \citep{pallottini2022,diMascia22} where we have a $\langle T_{\rm d} \rangle_{\rm M}$, but not a $\bar{T}_{\rm d}$, as the physical temperature of dust grains is derived from radiative equilibrium, and not drawn from our distribution. 
The standard deviation $\sigma_T$ and skewness $\zeta$ essentially inform us about the covering fraction of dense star-forming regions that host warm/hot dust pockets ($\gtrsim 60$ K, \citealt{sommovigo2020}), as well as the star formation efficiency within these regions compared to the diffuse ISM. 

\textit{In summary}: Our physical model for the multi-temperature SED of a galaxy is governed by four free parameters: three governing the dust temperature PDF ($\bar{T}_{\rm d}$; $\sigma_T$; $\zeta$), plus the total dust mass ($M_{\rm d}$) which acts as a normalization factor. Despite its simplicity, this model effectively reproduces the PDFs obtained from radiative-transfer post-processed high-resolution hydro simulations while allowing exploration of a much larger parameter space at negligible computational cost.

This is illustrated in Fig. \ref{fig:PDFandSED_SERRAlike}, where we compare the dust temperature PDF generated by our analytical model with that of Zinnia (a.k.a. serra05:s46:h0643), a typical\footnote{Zinnia is typical in terms of the properties under investigation here, i.e., the IR SED and dust temperature PDF.} SERRA galaxy \citep{pallottini2019,pallottini2022} located at $z=6.671$ \citep[see also][]{sommovigo2021}. Details on the simulation setup and radiative transfer post-processing with SKIRT \citep{camps2015} are provided in \cite{behrens2018,pallottini2022}. 
Using parameters \(\bar{T}_{\rm d} = 60\ \mathrm{K}\), \(\sigma_T = 16\ \mathrm{K}\), and \(\zeta = 0.98\), our model successfully reproduces the simulated \(T_{\rm d}\) PDF (both mass-weighted and luminosity-weighted). This comparison demonstrates the ability of our analytical approach to approximate the complexity of a physical multi-temperature dust distribution. We discuss this in further detail in Section \ref{sec:discussion}, where we additionally include a comparison to the dust SEDs of observed high-redshift galaxies and the temperature PDFs of a simulated quasar host galaxy.

We produce the IR SED as the sum of a series of optically thin modified blackbodies (MBBs) weighted according to our PDF as:
\begin{equation}
\label{eq:MBB}
    F_{\rm \nu} =  g(z) M_{\rm d} \kappa_{\nu} \int [B_{\nu}(T_{\rm d}')-B_{\nu}(T_{\rm CMB})] f(T'_{\rm d}) dT'_{\rm d},
\end{equation}
where 
at wavelengths $\lambda > 20\,\mathrm{\mu m}$, $\kappa_{\rm \nu}$ can be approximated as \citep{draine2003}
\begin{equation}\label{opacity}
\kappa_{\nu} = \kappa_0 \left(\frac{\nu}{\nu_{0}}\right)^{\beta_d}.
\end{equation} 
The choice of $(\kappa_0, \nu_0, \beta_d)$ depends on the assumed dust properties. We consider Milky Way-like dust, for which standard values from theoretical models are $(\kappa_0, \nu_0, \beta_{\rm d})$ = (10.41 ${\rm cm^2 g^{-1}}$, $1900\, {\rm GHz}$, 2.03), see \cite{draine2003}. We discuss the impact of the assumed dust model in Section \ref{sec:discussion_Mdust_uncertainty}.

In the right panel of Fig.~\ref{fig:PDFandSED_SERRAlike} we compare the IR continuum SED produced by our model with that of the simulation. We can see that the two are remarkably similar, with the primary discrepancy being the absence of PAH features in our analytical model. These features, which in this case contribute \(\sim 15\%\) to the total IR luminosity, are well below the typical uncertainties on the IR luminosities of high-redshift galaxies -- as we will see in Section \ref{sec:results}. Moreover, PAH features for \(z \sim 7\) galaxies fall within the observational gap between ALMA and JWST, making them less critical for our analysis. We will revisit this in a follow-up work where we include simulated constraints from the proposed FIR probe \textit{PRIMA} \citep{moullet2023}. \\

In the following, we detach the analysis from direct comparisons with simulations, shifting focus toward exploring a broad parameter space to assess the capabilities of our analytical model. This approach allows us to systematically investigate the impact of varying dust temperature parameters—mean (\(\bar{T}_{\rm d}\)), variance (\(\sigma_T\)), and skewness (\(\zeta\))—on the resulting physical properties and observable quantities of galaxies, without having to run expensive simulations whose output in the dust temperature parameter is not as directly controllable. 

As shown in Fig.~\ref{fig:Expl_Param}, we analyze how these parameters affect the dust temperature PDFs (left panels), IR continuum SEDs (central panels), and cumulative dust mass distributions (right panels). The mean dust temperature, $\bar{T}_{\rm d}$, shifts the entire cumulative mass distribution, resulting in variations in the luminosity-weighted temperature $\langle T_{\rm d} \rangle_\mathrm{L}$ that are directly comparable to changes in $\bar{T}_{\rm d}$ itself (i.e., the mass-weighted temperature). In contrast, increasing \(\sigma_T\) broadens the temperature distribution and primarily impacts the high-temperature end of the cumulative mass function, consistent with an increased contribution from hot dust pockets. This effect is most pronounced in the MIR portion of the SED, which is only weakly traced by ALMA at the considered redshift $z=7$ (in bands 9 and 10). Lastly, variations in the skewness parameter, \(\zeta\), introduce asymmetry into the PDF, with a higher skewness producing more pronounced hot dust tails. While this has only a minor effect on the overall SED shape, it reduces $\langle T_{\rm d} \rangle_\mathrm{L}$ for less skewed distributions, as the contribution from hotter dust components diminishes. Notably, the high-skewness scenario better reproduces the simulated PDFs.

In summary, the IR continuum SEDs (central panels) reflect how variations in $\bar{T}_{\rm d}$, \(\sigma_T\), and \(\zeta\) manifest observationally. $\bar{T}_{\rm d}$ shifts the peak of the SED to shorter wavelengths as it increases, while \(\sigma_T\) significantly alters the MIR emission. Skewness, on the other hand, has only a limited effect on the shape of the dust SED.

In the following, we compare the physical dust parameters (luminosity- and mass-weighted dust temperatures, $\langle T_{\rm d} \rangle_\mathrm{L}$ and $\langle T_{\rm d} \rangle_\mathrm{M}$, total IR luminosity \(L_{\rm IR}\), and dust mass \(M_{\rm d}\)) derived from the multi-temperature model with those recovered from simplistic single-temperature SED fitting. We adopt the same fitting procedure that would be applied to an actual ALMA observation, aiming to identify pathological areas of the parameter space where the commonly used single-temperature approximation leads to significant misinterpretation 
of the aforementioned dust properties. These results will be used to i) enforce a more accurate error estimation for traditional SED fitting procedures, and ii) identify (classes of) galaxies where our understanding of dust budgets and obscured SFRs is limited due to the lack of rest-frame MIR data.

\subsection{Constructing Mock ALMA Observations}
\label{sec:methods_construct_mockALMA}


To mimic what an observer would see, we sample the constructed dust SEDs in a range of possible ALMA bands. Our fiducial set of mock observations is drawn from Bands 6, 7 and 8, which are commonly used at $z\sim6-8$ as they cover the bright \ciil{} and/or \oiiil{} lines at these epochs (e.g., \citealt{hashimoto2019,harikane2020,sugahara2021,witstok2022,algera2024,mitsuhashi2024}). We assign each of these bands an $\mathrm{S/N} = 10$, which is achievable in reasonable observing times for moderately bright galaxies at this epoch. We refer to this realistic three-band setup as `Basic ALMA' throughout.

We also adopt a wider set of bands to mimic a very optimistic case of what may (soon) be achievable with ALMA at $z\sim7$. In this optimistic model, any systematic errors will be more apparent, as they are expected to dominate over the measurement uncertainties. We adopt each of Bands $3 - 8$, and assign them an $\mathrm{S/N}=10$, while also including Bands 9 and 10 with an $\mathrm{S/N} = 5$. The lower S/N for the higher-frequency bands is motivated by the fact that the uncertainty on the absolute flux calibration at these frequencies is approximately $20\,\%$.\footnote{\url{https://almascience.eso.org/documents-and-tools/cycle11/alma-technical-handbook}} While clearly an optimistic setup, we emphasize that the $z=7.31$ galaxy REBELS-25 has now been targeted in 
six distinct ALMA bands \citep{algera2024b}, with similarly expansive multi-wavelength follow-up now being obtained for other $z\gtrsim5$ galaxies \citep[e.g.,][]{villanueva2024}. Moreover, the Wideband Sensitivity Upgrade (WSU) scheduled for $2030$ is set to enhance the continuum sensitivity of ALMA by a factor of $\sim2-3\times$ \citep{carpenter2022}, making this multi-band coverage realistic for larger samples of (fainter) galaxies in the near future. We refer to this optimistic setup as `Super ALMA'. 

We note that in our analysis we adopt a fixed dust mass of $M_{\rm d}=10^7\,M_\odot$ for the mock galaxies, although this simply serves as a normalization factor. Since we are interested in relative quantities, such as the accuracy with which the input dust mass can be recovered [i.e., $\Delta \log(M_{\rm d}) = \log(M_\mathrm{d,fit} / M_\mathrm{d,input})$], our results do not depend on the overall dust mass, but rather on the assigned S/N of the observations, here assumed to be independent of $M_{\rm d}$.

\subsection{Fitting the Mock Observations}\label{sec:methods_fit_mockALMA}

We fit the global dust SEDs of our mock galaxies using a Monte Carlo Markov Chain (MCMC)-based fitting framework. For a complete description we refer the interested reader to \citet{algera2024}, where it was used to fit the SEDs of several high-redshift ($z\gtrsim5$) galaxies with multi-band ALMA photometry. In brief, an optically-thin, single-temperature MBB is assumed, and the \textit{emcee} library \citep{foreman-mackey2013} is used to explore the parameter space for $T_{\rm d,MBB}$, $M_{\rm d}$, and optionally $\beta_\mathrm{d}$. The CMB is corrected for in the fitting following \citet{dacunha2013}. 

When fitting the dust SEDs of galaxies with limited photometric sampling, as in the case of the Basic ALMA setup, carefully choosing one's priors is important (c.f., the discussion in \citealt{algera2024,algera2024b}). In this work, we adopt the same prior on the dust temperature and mass as used in several recent high-redshift studies (e.g., \citealt{algera2024,algera2024b,bakx2024,chen2024}; Van Leeuwen et al.\ submitted), to ensure consistency with the approach a typical observer would take. We adopt a wide, flat prior on the dust temperature, ranging from the CMB temperature [$T_\mathrm{CMB} = 2.73 \times (1 + z)\,\mathrm{K}$] to an upper limit of $150\,$K, after which the prior is smoothly decreased by a Gaussian with standard deviation $\sigma = 30\,$K to avoid fitted dust temperatures from increasing indefinitely when the data are not sufficiently constraining. We also adopt a flat prior on the dust mass from $\log(M_{\rm d} / M_\odot) \in [4, 12]$, though note that the precise range adopted does not affect our results.

For the dust emissivity index $\beta_\mathrm{d}$, we explore two scenarios, again following what an observer would typically do. In the Basic ALMA setup where the Rayleigh-Jeans tail is not accurately sampled, we adopt a fixed $\beta_\mathrm{d} = 2.0$ in the fit (following e.g., \citealt{sugahara2021,witstok2022,algera2024}). While it is also possible to adopt a Gaussian prior on $\beta_\mathrm{d}$ in order to marginalize across the uncertainty on this parameter, this tends to skew the recovered $\beta_\mathrm{d}$ to shallower values (e.g., \citealt{witstok2023_beta,algera2024,mitsuhashi2024}). In our Super-ALMA setup, we do allow for variation in $\beta_\mathrm{d}$ in the fit, as in this scenario the Rayleigh-Jeans tail is accurately sampled by observations in ALMA Bands 3 and 4 (c.f., \citealt{bendo2025}). We follow \citet{algera2024b} by adopting a flat prior on the emissivity index of $\beta_\mathrm{d} = [1.0, 4.0]$. 

From the fits we obtain full posterior distributions for $T_{\rm d,MBB}$, $M_{\rm d}$ and -- if varied -- $\beta_\mathrm{d}$. For each parameter, we quote the median value, with the $16-84^\mathrm{th}$ percentile spread representing the uncertainty (equivalent to $1\sigma$). Infrared luminosities $L_\mathrm{IR}$ and their corresponding uncertainties are calculated by integrating MBBs computed from a representative subset of the posterior parameter values across rest-frame $8-1000\,\mu\mathrm{m}$, utilizing the dust temperature corrected for CMB heating (i.e., the inverse of Equation \ref{eq:Tcmb}).

\section{Results}\label{sec:results}
Using the machinery outlined in the previous section, we construct mock dust SEDs across a grid of $\bar{T}_{\rm d}$, $\sigma_T$ and $\zeta$. For the dust temperature, we explore a range of $\bar{T}_{\rm d} = 30 - 100\,\mathrm{K}$ in steps of $\Delta \bar{T}_{\rm d} = 10\,\mathrm{K}$, while for $\sigma_{T}$ we adopt a range of $\sigma_{T} = 0.1 - 0.6\,\bar{T}_{\rm d}$ in steps of $0.1$. Finally, for the skewness we adopt two fixed values of $\zeta \in \left\{0.01, 0.98\right\}$, given its limited effect on the overall shape of the dust SED.

\begin{figure*}
    \centering
    \includegraphics[width=0.9\textwidth]{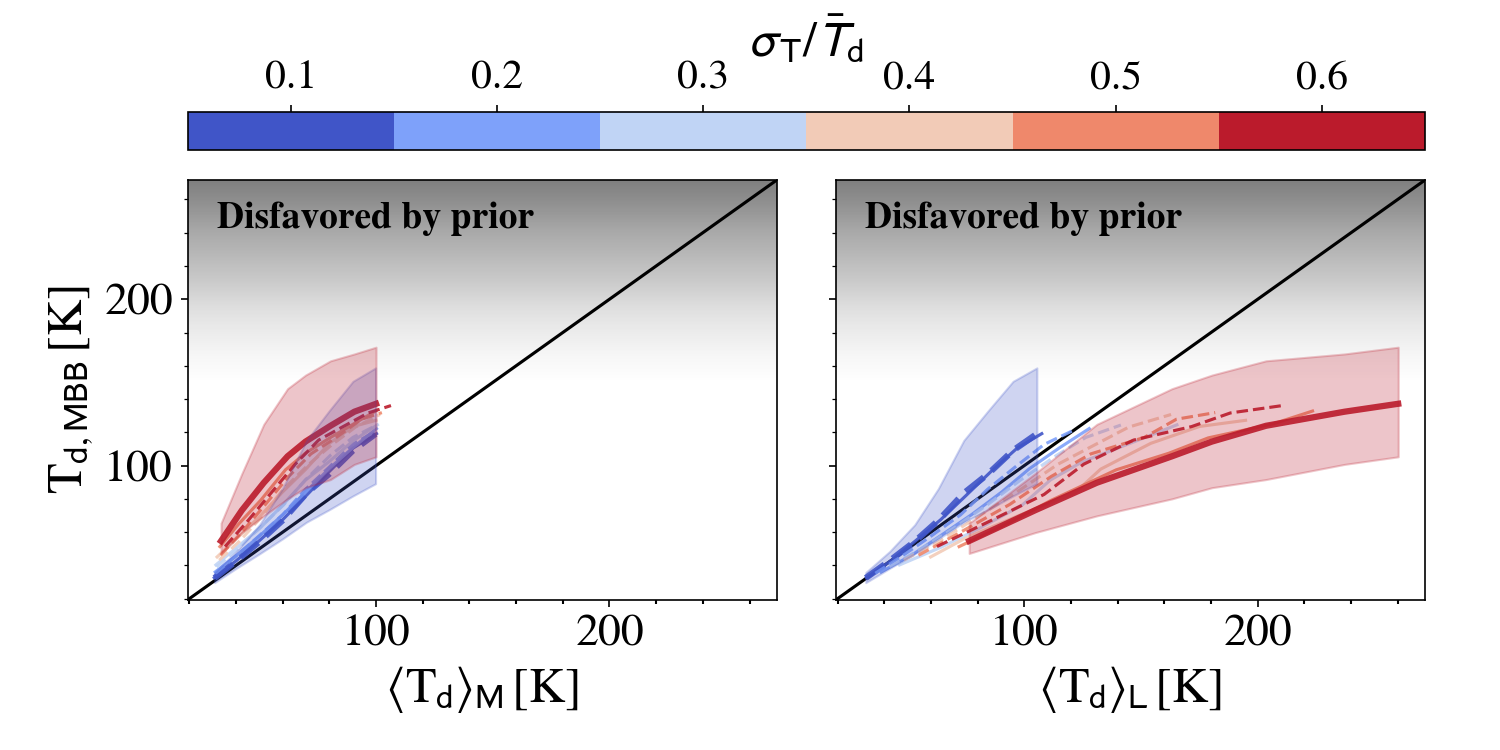}
    \includegraphics[width=0.9\textwidth]{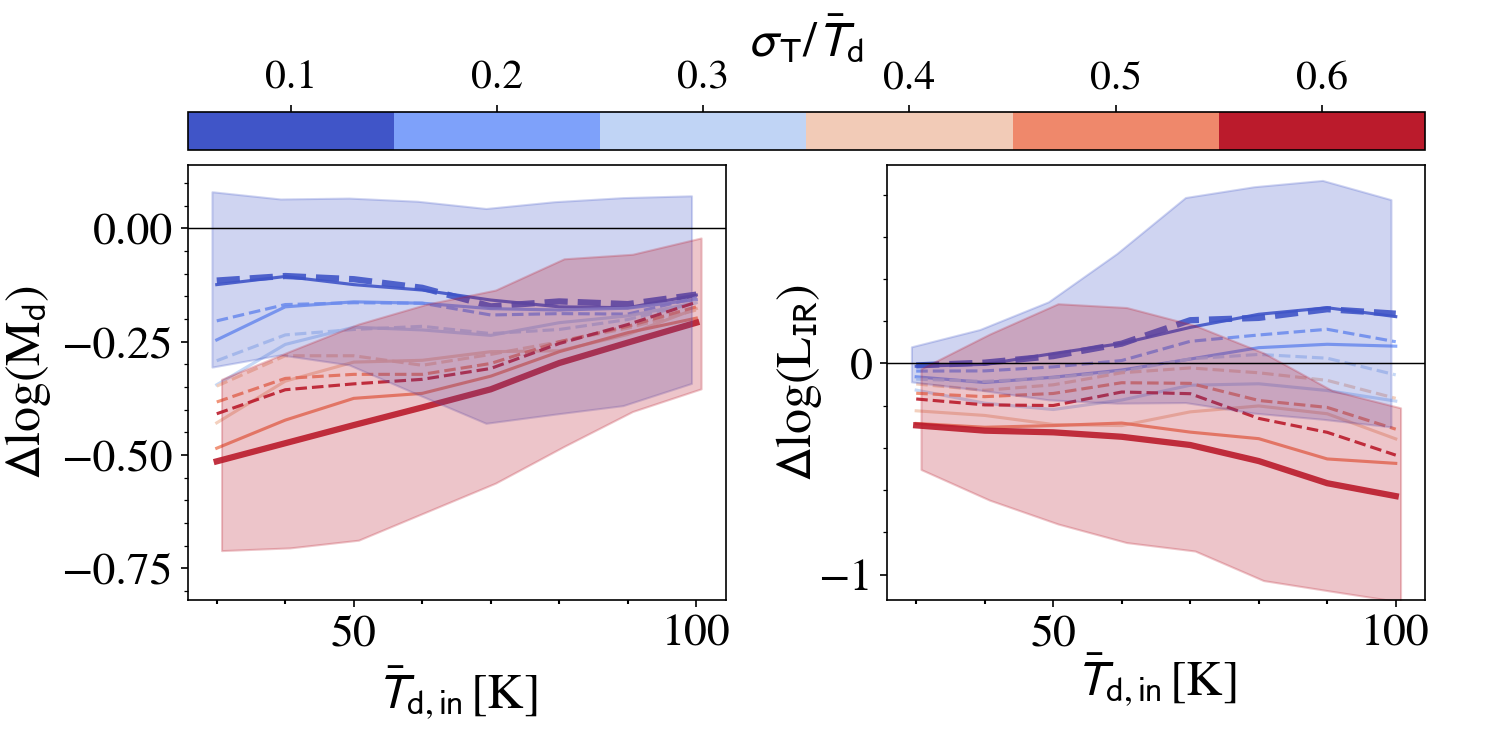}
    \caption{\textbf{The effect of multi-temperature dust on the recovered dust temperature, mass and IR luminosity, assuming the `Basic ALMA' setup with S/N=10 detections in ALMA Bands 6, 7 and 8.} \textit{Top row}: fitted MBB dust temperature as a function of the input mass-weighted temperature (left) and luminosity-weighted temperature (right). The color-coding represents the width of the dust temperature distribution, ranging from narrow (blue) to wide (red). Solid and dashed lines correspond to a skewness of $\zeta=0.01$ and $0.98$, respectively. Shaded regions represent the error on the fitted temperature, and are shown only for the narrowest and widest dust temperature distribution for clarity (thicker lines). The mass-weighted temperature tends to be overestimated when the dust temperature PDF becomes broad, while the luminosity-weighted temperature is instead underestimated. \textit{Bottom row}: accuracy with which the dust mass [left; defined as $\log(M_\mathrm{d}) = \log(M_\mathrm{d,fit}/M_\mathrm{d,in})$] and IR luminosity (right; defined analogously) can be recovered, as a function of the input mass-weighted temperature. For broad $T_{\rm d}$ distributions, dust masses and IR luminosities can both be underestimated by up to $\sim0.5\,\mathrm{dex}$. However, for typical mass-weighted temperatures ($\bar{T}_{\rm d} \lesssim 50\,\mathrm{K}$), the IR luminosity is typically well-recovered, while in this regime observed dust masses will be more systematically off.}
    \label{fig:basicALMA}
\end{figure*}

\subsection{Basic ALMA}
\label{sec:results_basicALMA}

\begin{figure*}
    \centering
    \includegraphics[width=0.9\textwidth]{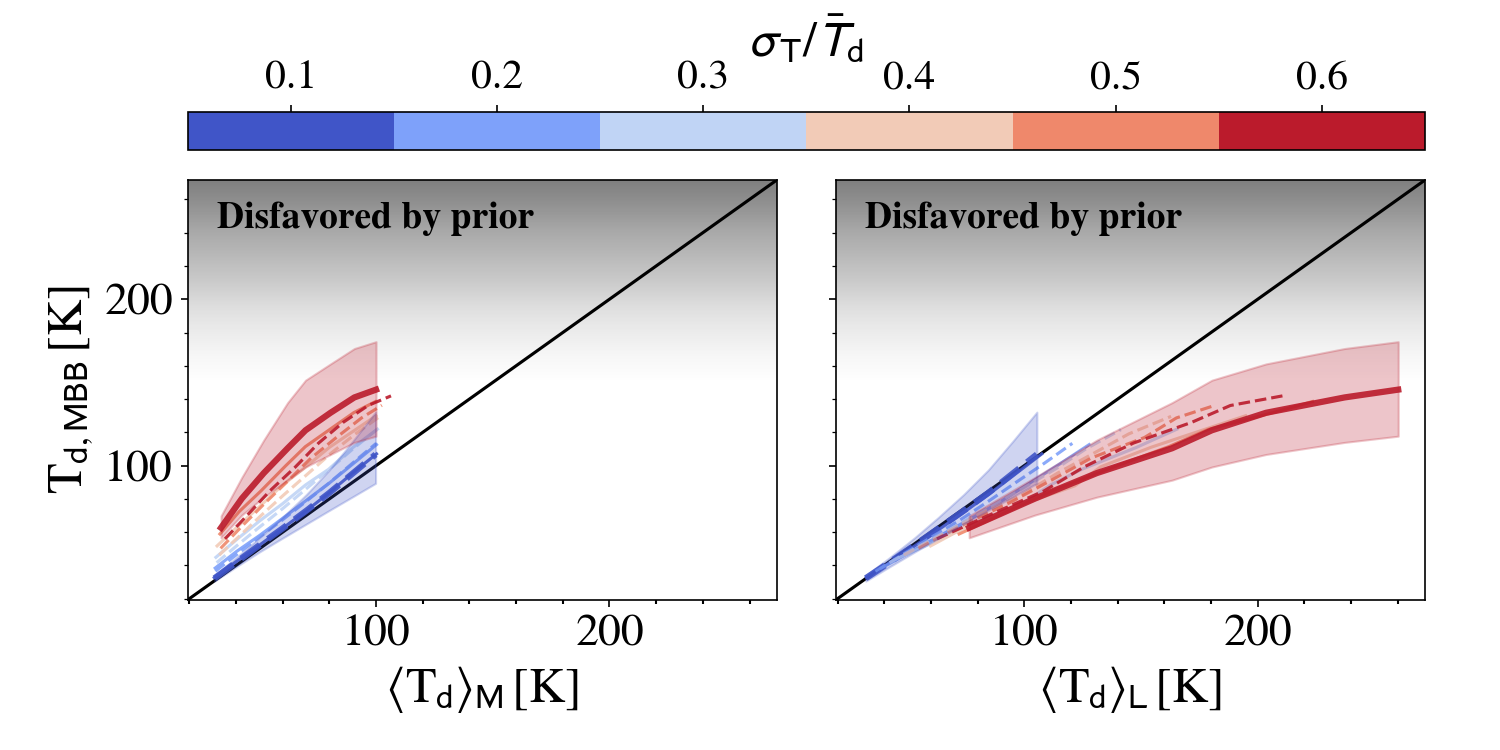}
    \includegraphics[width=0.9\textwidth]{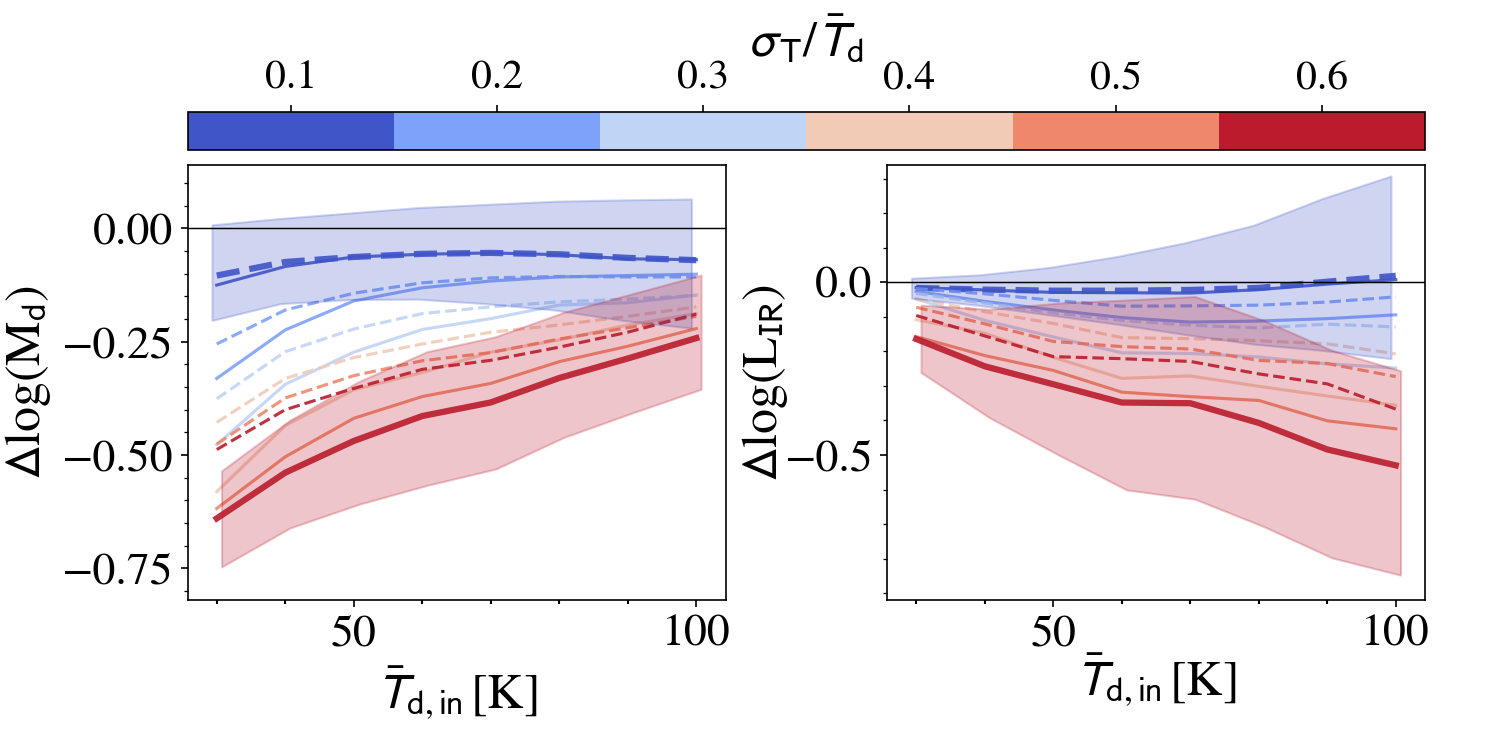}
    \caption{Same as Fig.\ \ref{fig:basicALMA}, now using the `Super ALMA' setup which covers the dust SED in all of Bands 3 - 10 and leaves $\beta_\mathrm{d}$ as a free parameter in the fit (see text). Unlike the `Basic ALMA' setup, systematic errors related to broad multi-temperature dust distributions now start to dominate, and are not accurately captured within the observational uncertainties.}
    \label{fig:superALMA}
\end{figure*}

We first fit the full grid of constructed dust SEDs using the `Basic ALMA' mock observations as introduced in Section \ref{sec:methods_construct_mockALMA}, and recall that this analysis adopts a fixed $\beta_\mathrm{d} = 2.0$. We show the fitted dust temperature $T_\mathrm{d,MBB}$ as a function of the input mass- and luminosity-weighted dust temperatures in the upper panel of Figure \ref{fig:basicALMA}. For each subplot, we show the full grid of $\sigma_{T}$ via the color-coding, while fits to input SEDs with $\zeta = 0.01$ ($\zeta = 0.98$) are shown via a dashed (solid) line. For the sake of readability, we show only the uncertainties on the fitted temperature for the scenarios with ($\sigma_T, \zeta$) = ($0.1, 0.01$) and ($\sigma_T, \zeta$) = (0.6, 0.98), i.e., the dust SEDs with the narrowest and widest temperature distribution, respectively. 

As expected, both the mass- and luminosity-weighted temperatures are most accurately recovered when the width of the input temperature distribution is narrow. For larger values of $\sigma_T$, $\langle T_\mathrm{d}\rangle_\mathrm{M}$ and $\langle T_\mathrm{d}\rangle_\mathrm{L}$ naturally start to diverge, with the former being overestimated, while the latter is instead underestimated. Indeed, the MBB dust temperature tends to fall in between the mass- and luminosity-weighted temperatures. We note that at high input temperatures, the adopted prior on the fitted dust temperature starts affecting the results; this is most clearly visible as a flattening of the fitted MBB dust temperature for high luminosity-weighted temperatures, as the prior prevents the fitted $T_{\rm d,MBB}$ from increasing indefinitely. This flattening occurs for high mass-weighted temperatures as well, albeit to a lesser extent as this temperature is always lower than the luminosity-weighted one. 

In the bottom panel of Figure \ref{fig:basicALMA}, we compare the recovered dust masses and IR luminosities with the input values. Because the mass-weighted temperature tends to be \textit{overestimated} by the single-temperature model, the dust masses are slightly \textit{underestimated}. This effect is mostly negligible for narrow dust temperature PDFs ($\sigma_T / \bar{T}_{\rm d}\sim 0.1$), as the systematic offset with respect to the input dust mass is well within the measurement uncertainties. However, for wide dust temperature PDFs, the effect can be substantial, with dust masses being overestimated by $\sim0.5\,\mathrm{dex}$, which is larger than the measurement error. This effect is strongest for low $\bar{T}_{\rm d}$, as for higher temperatures the imposed prior prevents the MBB temperature from strongly overestimating the mass-weighted one.

Unlike for the dust masses, however, the IR luminosities are typically rather well recovered by the fit. For narrow $\bar{T}_{\rm d}$ distributions, the luminosity-weighted and MBB-temperatures are in good agreement, which means that the IR luminosity can accurately be measured. Only for the widest dust temperature distributions and hottest input temperatures do we find $L_\mathrm{IR}$  to be slightly underestimated; this is mostly due to the luminosity-weighted temperature being underestimated by the fit, in part due to the impact of the adopted prior keeping the MBB temperature from increasing indefinitely. \\

We note that we adopt a fixed $\beta_\mathrm{d} = 2.0$ in our analysis, both a priori to generate the mock dust SEDs, and as a fixed value in our fits. In practice, there is a non-negligible scatter in $\beta_\mathrm{d}$ (e.g., \citealt{dacunha2021,bendo2023,witstok2023_beta,algera2024b,liao2024,tripodi2024,chiang2025}), which could point to variation in the intrinsic dust properties of high-redshift galaxies. However, the observed variation in $\beta_\mathrm{d}$ can also be due to fitting degeneracies (e.g., \citealt{juvela2013}), and multi-temperature dust (e.g., \citealt{shetty2009}; see also Section \ref{sec:discussion}). Regardless, the a priori choice of the `correct' $\beta_\mathrm{d}$ in our fitting procedure suggests that the systematic uncertainties in Figure \ref{fig:basicALMA} are likely being underestimated.

While a full exploration of the impact of the assumed $\beta_\mathrm{d}$ is beyond the scope of our work, we briefly investigate how our results would change had we adopted a fixed $\beta_\mathrm{d} = 1.5$ in our fits, while maintaining $\beta_\mathrm{d} = 2.0$ as our input value to construct the dust SEDs. In this case, we find that the fitted MBB temperature overestimates \textit{both} the mass- and luminosity-weighted temperatures, except for very high values of $\bar{T}_{\rm d}$ where the prior starts impacting the results. In practice, this means that dust masses are underestimated, typically by a systematic value of $\Delta\log(M_{\rm d}) \sim 0.5\,\mathrm{dex}$, but potentially by as much as $\sim1\,\mathrm{dex}$. 

For the IR luminosity, the results are not as catastrophic; while the luminosity-weighted temperature is overestimated by the fit, the fact that the dust mass is being underestimated partially compensates for this effect (recall that $L_\mathrm{IR} \propto M_{\rm d} T_{\rm d}^{4+\beta_\mathrm{d}}$). Nevertheless, for narrow dust temperature distributions with warm dust ($\bar{T}_{\rm d} \gtrsim 50\,\mathrm{K}$), the IR luminosity is typically overestimated by $\sim0.5\,\mathrm{dex}$. On the other hand, for wider distributions, the inferred IR luminosity is not systematically off unless the input mass-weighted temperature is sufficiently hot ($\bar{T}_{\rm d} \gtrsim80\,\mathrm{K}$), at which point $L_\mathrm{IR}$ starts to become underestimated by up to $\sim0.5\,\mathrm{dex}$ (in part due to the assumed prior).
\subsection{Super ALMA}
\label{sec:results_superALMA}

\begin{figure*}
    \centering
    \includegraphics[width=0.85\textwidth]{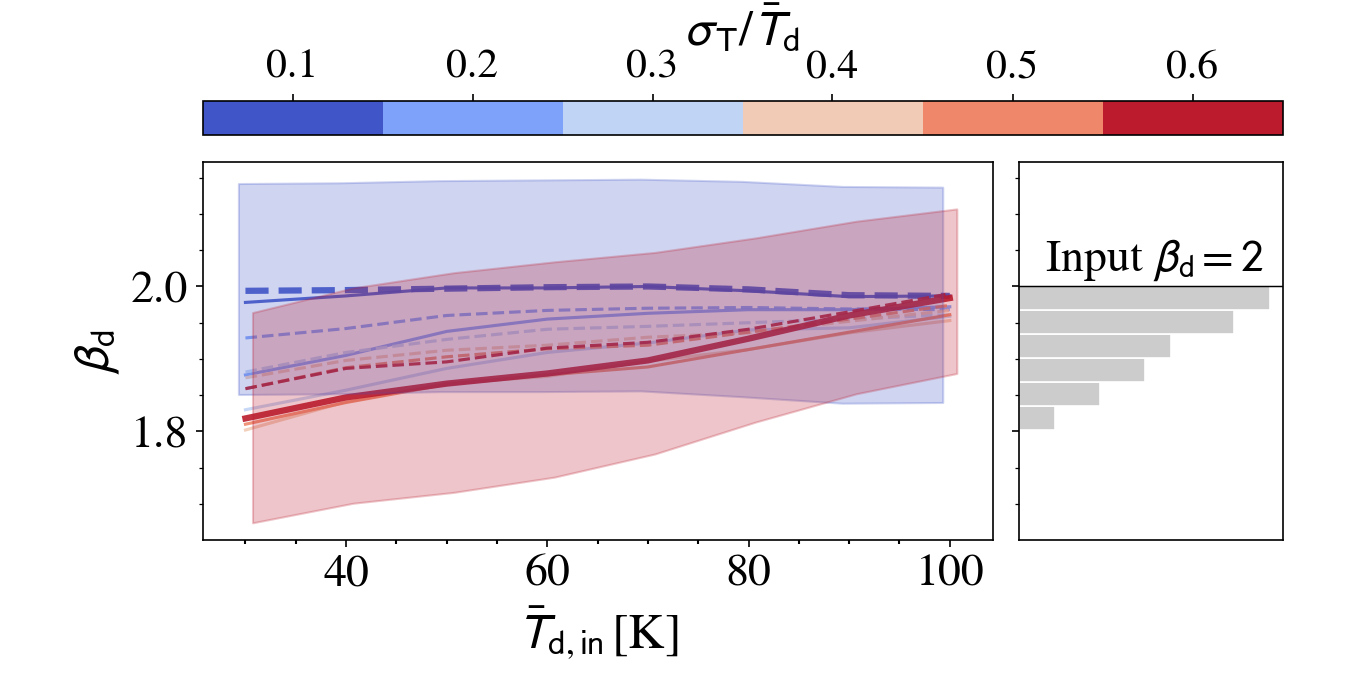}
    \caption{\textbf{Dust emissivity indices are biased to shallower values due to multi-temperature dust.} Recovered dust emissivity indices $\beta_\mathrm{d}$ for the Super ALMA setup, as a function of the input mass-weighted temperature (\textit{left}; symbols and line styles are the same as in Fig.\ \ref{fig:basicALMA}), and as a histogram combining all fits (\textit{right}). While the input value is fixed to $\beta_\mathrm{d} = 2.0$ for all mock dust SEDs, recovered dust emissivity indices end up being slightly shallower; this demonstrates that multi-temperature dust tends to flatten the observed emissivity index. The effect is most pronounced for wide temperature distributions with cold mass-weighted temperatures.}
    \label{fig:beta_hist}
\end{figure*}

We proceed by fitting our mock MBBs with the `Super ALMA' setup, which we recall utilizes observations in all of Bands 3 - 10. With this expanded wavelength coverage, we can 1) investigate whether limitations in single-temperature models systematically bias inferred dust masses and IR luminosities well beyond the measurement errors, and 2) assess whether the recovered dust emissivity index $\beta_\mathrm{d}$ is biased in single-$T_{\rm d}$ models.

We show our results in Figure \ref{fig:superALMA}. Similar to what we found for the `Basic ALMA' setup, the dust temperature inferred from MBB fitting does not correspond to any physical temperature (i.e., neither $\langle T_{\rm d} \rangle_\mathrm{M}$ nor $\langle T_{\rm d} \rangle_\mathrm{L}$) when the underlying multi-temperature distribution is wide. 

Despite the expanded ALMA coverage, we find that dust masses remain systematically underestimated for wide temperature distributions, up to $\sim0.6\,\mathrm{dex}$ for cold mass-weighted temperatures ($\bar{T}_{\rm d}\lesssim40\,\mathrm{K}$). Unlike for the Basic ALMA setup, these systematic uncertainties are now much larger than the typical measurement errors on the dust mass of $\pm0.1-0.2\,\mathrm{dex}$. As before, however, infrared luminosities can be recovered to within $\lesssim0.3\,\mathrm{dex}$ for all but the widest and warmest temperature distributions where $L_\mathrm{IR}$ may be underestimated by up to $\sim0.5\,\mathrm{dex}$. However, these extreme scenarios require mass-weighted dust temperatures well above $\bar{T}_{\rm d} \gtrsim 60\,\mathrm{K}$, which we find in Section \ref{sec:discussion} are unlikely even for powerful $z\sim6$ quasars \citep[see also][]{diMascia22}. Overall, we thus find that, for $\bar{T}_{\rm d} \lesssim 50\,\mathrm{K}$, infrared luminosities can generally be recovered to within $\lesssim0.3\,\mathrm{dex}$.

We next turn to the dust emissivity index, which has lately garnered some attention at high redshift as -- in addition to attenuation curves at UV/optical wavelengths (e.g., \citealt{witstok2023_2175,markov2024,sanders2024,fisher2025}) -- it may provide direct insight into the physical properties of, and ISM conditions surrounding, astrophysical dust grains (e.g., \citealt{kohler2015,ysard2019,hirashita2023,algera2024b}). 

We show the distribution of recovered dust emissivity indices in Figure \ref{fig:beta_hist}, recalling that our input value is fixed to $\beta_\mathrm{d} = 2.0$. While the median recovered value ($\beta_\mathrm{d} \approx 1.94$) is similar to the input one, the distribution is truncated at $\beta_\mathrm{d} < 2$. In other words, only dust emissivity indices shallower than the input value are recovered, which agrees with previous works suggesting $\beta_\mathrm{d}$ flattens due to so-called "temperature mixing" \citep{hunt2015}. The largest differences between the recovered and input $\beta_\mathrm{d}$ are found for cold mass-weighted temperatures $\bar{T}_\mathrm{d}$ with a wide distribution ($\sigma_T \gtrsim 0.3 \bar{T}_\mathrm{d}$). We discuss this in further detail in the next section.

\section{Discussion}\label{sec:discussion}

In the following, we place our model in the context of simulations of high-redshift astrophysical sources, as well as observed galaxies with multiple FIR continuum measurements. Finally, we compare the uncertainties in galaxy dust masses resulting from multi-temperature dust to other sources of (systematic) error.

\begin{figure*}
    \centering
    \includegraphics[width=0.9\textwidth]{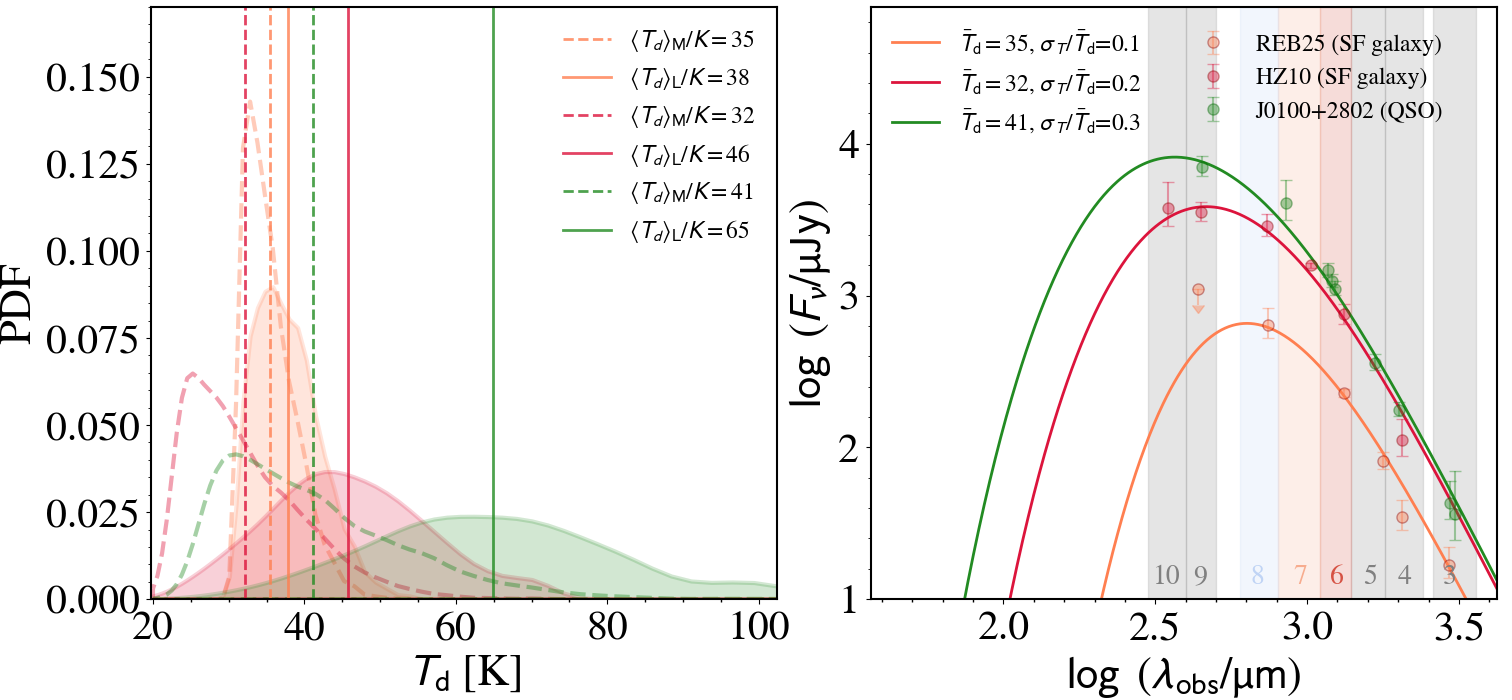}
    \includegraphics[width=0.9\textwidth]{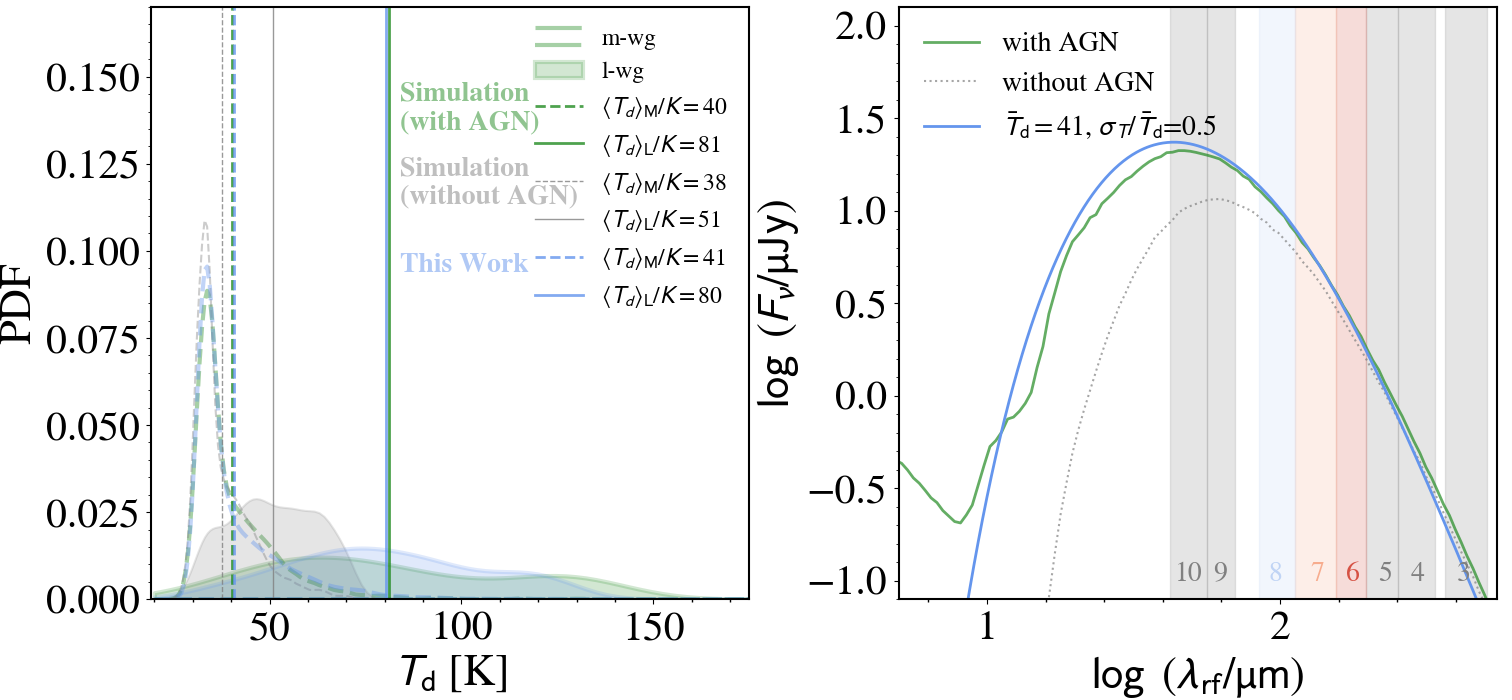}
    \caption{\textbf{Our simple analytical model can reproduce the complex IR SEDs of high-redshift galaxies and quasars.} Similar structure to Fig.~\ref{fig:PDFandSED_SERRAlike}. The \textit{top panels} show two high-redshift galaxies (HZ10 and REBELS-25 at $z=5.7$ and $z=7.3$ respectively) and a $z=6.3$ quasar (J0100+2802), all exhibiting different dust and global properties (see text). Since the temperature PDF is not available for the observed sources, we illustrate how our model can reproduce the FIR continuum data. We find that for both the star-forming (SF) source HZ10 and, even more so, the QSO \citep{tripodi2023}, the best fit requires a larger $\sigma_T / \bar{T}_{\rm d}$ (0.2 and 0.3, respectively). This is consistent with the presence of a warm dust component, which increases the luminosity-weighted dust temperature beyond the mass-weighted temperature and the temperature inferred from SED fitting.
    The \textit{bottom panels} compare our model to the zoom-in simulations from \citet{diMascia22}, specifically to the faint AGN (B in \citealt{diMascia22}) at $z = 6.1$. Our model (in blue) successfully reproduces both the temperature PDF and SED of the AGN while remaining agnostic about the source of dust heating. In this case, we also compare with the simulated temperature PDF derived from radiative transfer post-processed simulations, both with (green) and without (grey) AGN contribution. The inclusion of AGN activity introduces a warm dust component, which minimally affects the mass-weighted temperature PDF but significantly shifts the luminosity-weighted distribution, boosting the emission at $\lambda_\mathrm{rf} < 100 \, \mu$m. This emphasizes the importance of shorter-wavelength ALMA bands (shortwards of ALMA Band 8) for properly characterizing the dust properties of high-redshift galaxies.}
    \label{fig:comparison_sources_discussion}
\end{figure*}

\subsection{Multi-temperature dust across the galaxy (and AGN) population}

\begin{figure}
    \centering
    \includegraphics[width=0.45\textwidth]{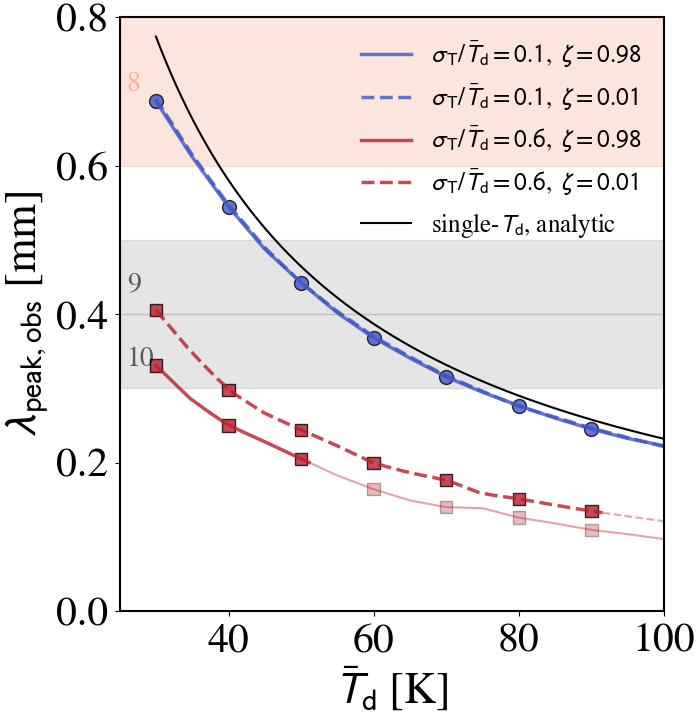}
    \caption{\textbf{Multi-temperature dust can shift the peak of the dust SED well beyond the ALMA observing bands, even when the mass-weighted temperature is low.} We show the peak wavelength $\lambda_{\rm peak,obs}$ of a hypothetical $z=7$ galaxy as a function of the mass-weighted dust temperature, and overlay the high-frequency ALMA bands (8, 9, 10). The red (blue) lines correspond to the widest (narrowest) dust temperature PDFs explored in our model. The onset of the higher transparency of the line/symbols corresponds to the mean temperature $\bar{T}_{\rm d}$ at which the fitted $L_{\rm IR}$ deviates by more than $>0.3\,\mathrm{dex}$ from the true value.}
    \label{fig:lambdaPeak}
\end{figure}

Our multi-temperature dust model is capable of reproducing a broad range of dust temperature distributions, which can characterize different observed high-redshift systems from quasars \citep[e.g.,][]{Priddey2021,decarli2023,tripodi2023,tripodi2024,fernandez-aranda2025}, to -- possibly -- the elusive little red dots (LRDs, \citealt{Matthee24}), to evolved, metal-rich systems like some of the REBELS galaxies \citep{algera2024,rowland2024,rowland25}. This is shown in the upper panel of Fig.~\ref{fig:comparison_sources_discussion}, where we show three example sources -- two star-forming galaxies and a quasar. We show the $z=5.657$ galaxy HZ10 (\citealt{capak2015,faisst2020b,villanueva2024}; Algera \& Herrera-Camus et al.\ in preparation), the $z=6.327$ quasar J0100+2802 \citep{tripodi2023}, and the $z=7.31$ galaxy REBELS-25 \citep{inami2022,hygate2023,algera2024,algera2024b,rowland2024}.

At $z\sim 4-7$, we expect higher mass-weighted dust temperatures $\left<T_{\rm d}\right>_{\rm M}=40-60\, \mathrm{K}$ in starbursting systems such as bright \oiiil{} emitters (e.g., \citealt{bakx2020,jones2024,vallini2024}), due to their hard radiation fields. Indeed, as predicted by the analytical model from \citet{sommovigo2022}, in a simple single-dust temperature and single-phase ISM approximation, the dust temperature ($T_{\rm d,S22}$) is expected to depend on galaxy parameters as:
\begin{equation}\label{Tdcosm}
    T_{\rm d,S22} = 29.7 \left[ \frac{(1-e^{-\tau_{\rm eff}})}{Z}\left(\frac{\rm Gyr}{t_{\rm dep}}\right) \right]^{1/(4+\beta_{\rm d})} {\rm K}
\end{equation}
where $\beta_{\rm d}=2.03$ in the \cite{draine2003} SMC or MW physical dust model, $t_{\rm dep}$ is the total gas depletion time, $Z$ is the metallicity in solar units, and $\tau_{\rm eff}$ is the galaxy's effective optical depth ($\propto N_{\rm H}$; the gas column density). Therefore, galaxies that are young, metal-poor, having most of their dust still within parental clouds (i.e., $\tau_{\rm eff}\sim 1$) and/or undergoing a burst of star formation (i.e., a lower $t_{\rm dep}$) should host warm/hot dust. This is the case for the simulated galaxy Zinnia from the SERRA suite (shown in Fig. \ref{fig:PDFandSED_SERRAlike}), which as discussed in Appendix A of \cite{sommovigo2021}, deviates upwards from the Kennicutt-Schmidt relation (with its SFR surface density $\Sigma_{\rm SFR}$ being $\sim 4\times$ higher than expected based on its gas surface density). 
Conversely, we expect colder dust temperatures ($\left<T_\mathrm{d}\right>_\mathrm{M}\sim 30-35\,$K, see Fig.~\ref{fig:comparison_sources_discussion}) in more evolved galaxies where large amounts of dust and metals are accumulated -- such as REBELS galaxies where we measure high dust-to-metal ratios \citep{algera2025} -- and processed in the diffuse lower density ISM, further away from young strongly UV-emitting sources and at $\tau_{\rm eff}\ll 1$. \\

The skewness ($\zeta$) and width ($\sigma_{T}$) of the dust temperature distribution reflect the properties of the gas (column) density distribution across different regions of the ISM \citep{ChiaYu_2018,Martizzi_2019,pallottini2019,Bieri_2023,vallini2024}, as well as—on smaller scales—within individual star-forming regions \citep[see e.g.,][]{Federrath2012,Menon_2024}.
We expect that the width of the dust temperature distribution, parametrized by $\sigma_{T}$, correlates with galaxy age via its star formation history. In the early stages, galaxies form stars within dense, compact star-forming regions, leading to a narrow temperature PDF centered around the warm dust temperatures typical of high-$z$ GMCs ($60$–$100$ K). As the galaxy evolves and dust spreads into the diffuse ISM, lower temperatures are also reached \citep{Ferrara2017}, broadening the temperature PDF (with the temperature floor being set by the CMB temperature at $T_{\rm CMB}\gtrsim 20\,\mathrm{K}$ at $z \gtrsim 6$). This increase in $\sigma_{T}$ mirrors the broadening of the gas density distribution \citep{Menon_2024}, which arises due to growing turbulence—an effect that scales with the galaxy’s SFR (\citealt{Semenov2017}; see also \citealt{sommovigo2020} for cloud-scale analytical modeling). Finally, as galaxies eventually transition to a quiescent state, the dust temperature PDF narrows again, this time shifting to lower temperatures, ultimately settling into equilibrium with the CMB.

As our model is agnostic to the source of dust heating, it can also be used to describe dust heated by an AGN. Radiative Transfer (RT) post-processed simulations (e.g., \citealt{diMascia21,diMascia22}) suggest that AGN-heated dust can reach even more extreme temperature distributions, resembling our high-\(\sigma_T\) scenario (e.g., \(\sigma_T/\bar{T}_{\rm d} = 60\%\)). In the lower panel of Fig.~\ref{fig:comparison_sources_discussion}, we show the comparison to one of the zoom-in simulated galaxies from \cite{Valentini21}  whose RT post-processing is presented in \cite{diMascia22}. We show the simulation with and without AGN heating. We can see that turning on the AGN leaves the mass-weighted PDF with $\left<T_{\rm d}\right>_{\rm M}=40$ K essentially unchanged, but requires a much larger  \(\sigma_T/\bar{T}_{\rm d} = 0.5\) (and maximal skewness) to reproduce the resulting IR-bright hot dust component. Potentially even more extreme are the LRDs, for which \citet{setton2025} recently argued that, if they contain substantial amounts of dust, they need to have typical temperatures of $T_{\rm d, MBB} \sim 150 - 300\,\mathrm{K}$ to emit predominantly in the observational gap between \textit{JWST}/MIRI and ALMA. Such temperatures would be well above even the luminosity-weighted temperature for the \citet{diMascia22} quasar of $\langle T_{\rm d}\rangle_\mathrm{L} \sim 80\,\mathrm{K}$ (Fig.\ \ref{fig:comparison_sources_discussion}, and recall that $\langle T_{\rm d}\rangle_\mathrm{L} \gtrsim T_{\rm d, MBB}$; Fig.\ \ref{fig:basicALMA}).

Overall, the above comparisons highlight the potential of our model to quantify systematic errors in the inference of dust properties from observations in a wide range of astrophysical sources.
Crucially, we showed that independently from the mean temperature, large $\sigma_{T}$ values introduce significant errors, as increasing amounts of IR emission shift to wavelengths not traced by ALMA (see also \citealt{diMascia22} for a related discussion specifically for AGN hosts). This is exemplified in Fig.~\ref{fig:lambdaPeak} where we show the location of the peak of the IR emission as a function of different values of the mass-weighted temperature $\bar{T}_{\rm d}$ and of $\sigma_T$. As expected, the peak shifts outside of the observable range by ALMA for $\bar{T}_{\rm d}>70\ \mathrm{K}$ at $z=7$ for narrow $\sigma_T/\bar{T}_{\rm d}=0.1$ (consistently with the single-temperature analytical prediction showcased by the black line and discussed in \citealt{bakx2021}). However, it is interesting to note that for broad $\sigma_T/\bar{T}_{\rm d}=0.6$, the peak is located shortwards of ALMA band 10 even at $\bar{T}_{\rm d} \gtrsim 35\ \mathrm{K}$, resulting in an underestimation of the total IR luminosity of $\gtrsim0.3\,\mathrm{dex}$ already at temperatures $\bar{T}_{\rm d} \gtrsim50\,\mathrm{K}$, even for the optimistic Super-ALMA configuration. Moreover, the uncertainty on the IR luminosity increases with $\sigma_T$ at fixed $\bar{T}_\mathrm{d}$ (c.f., Fig.\ \ref{fig:superALMA}), which is a direct consequence of the peak of the SED shifting beyond the ALMA coverage. \\

Our analytical approach is complementary to previous works that investigate the biases introduced by using simple, optically thin, single-temperature modified blackbody models to fit the FIR SEDs of simulated high-redshift galaxies (e.g., \citealt{ma2019,shen2022,vijayan2022,lower2024}). We focus here in particular on \citet{lower2024}, who fit mock dust SEDs constructed from post-processed $z\approx6.5$ dusty galaxies in the Cosmic Sands simulations using a two-band setup roughly corresponding to ALMA Bands 6 and 8. They find that MBB-fitted dust temperatures generally do not correspond to either the mass- or luminosity-weighted temperature, which can lead to biases in recovered dust masses and IR luminosities. However, the large fitting uncertainties complicate the interpretation, as does the fact that \citet{lower2024} find the dustiest galaxies in their simulations to be optically thick in the far-infrared ($\tau > 1$ at rest-frame $\lambda > 100\,\mu\mathrm{m}$). For their optically thin subset, however, MBB dust temperatures generally exceed the mass-weighted temperature, while they underestimate the luminosity-weighted value -- similar to what we find in our analysis (Section \ref{sec:results}).

For the time being, in our analytical model, we assume optically thin dust and focus on how a more or less extended dust temperature distribution affects the inferred dust properties. Notably, even in systems that are optically thin in the infrared, very hot dust temperatures can still be achieved due to the strong contrast in absorption opacity between different wavelengths. Specifically, a galaxy that remains transparent at $\lambda = 100\ \mathrm{\mu m}$ can be fully obscured in the ultraviolet at $\lambda = 1600\ \angstrom$, as the opacity contrast is $2.4 \times 10^3$.
The impact of optically thick dust will be explored in future work, where we will also incorporate simulated constraints from \textit{PRIMA} \citep{moullet2023}. Indeed, for dust to become optically thick in the regime accessible to ALMA at $z=7$, i.e. at $\lambda>37.4\ \mathrm{ \mu m}$ for Super-ALMA and $\lambda>75\ \mathrm{ \mu m}$ for basic-ALMA, the required column densities according to the \cite{draine2003} models are $N_{\rm H} > 2 \times 10^{23}\ \mathrm{cm^{-2}}$ and $N_{\rm H} > 10^{24}\ \mathrm{cm^{-2}}$, respectively. Such high-density regions likely have low covering fractions -- and thus contribute subdominantly to the dust mass budget -- in the mostly UV-selected, and UV-bright galaxies observed at high-$z$. This may be different in sub-millimeter-selected galaxies such as the South Pole Telescope sources \citep{reuter2020}, and possibly the compact LRDs \citep{Perez-Gonzalez24}.  

A further complication arises when considering grain size evolution \citep[see e.g.,][]{Hirashita20,Hu_2019,Hu_2023,Choban2022,Choban2024,Choban2024b}, which directly affects $\beta_{\rm d}$ (currently fixed to the Milky Way value in our SED modeling) and influences the evolving dust temperature PDF. The grain size distribution and multi-temperature PDF are inherently degenerate in shaping the IR SED \citep{ysard2019}. We find that deviations in $\beta_{\rm d}$ due to a broad $\sigma_T$ tend to skew the inferred $\beta_\mathrm{d}$ (under the single-$T_{\rm d}$ assumption) toward lower values than the input $\beta_\mathrm{d} = 2$, consistent with previous analytical modeling \citep{shetty2009}, observations of local galaxies \citep{kirkpatrick2014,hunt2015}, and results from RT post-processed zoom-in hydrodynamic simulations \citep{behrens2018,pallottini2022}. Thus, if $\beta_{\rm d} > 2$ is robustly recovered from high-SNR, well-sampled continuum SEDs \citep[see e.g.,][]{algera2024b}, this could serve as an indication of grain size evolution.
Observations combining Herschel, IRAS, and Planck \citep{juvela2015} show that within our Galaxy, a higher $\beta_{\rm d} \approx 2.2$ is measured in cold, dense clumps, whereas the average global value is $\beta_{\rm d} \approx 1.84$. These dense environments are prime sites for grain growth (e.g., \citealt{hirashita2012}), suggesting a possible link between $\beta_{\rm d}$ and the grain size distribution. However, as highlighted by \citet{kohler2015}, this would be a secondary effect, arising from the dependence of key dust processes rates —such as grain growth \citep{hirashita2012,michalowski2015,mancini2015,ferrara2016} and sputtering \citep{Draine79,Jones1996ApJ,Tielens99}—on the surrounding gas density.

\subsection{Multi-temperature dust: implications for early dust production}

As shown in the previous sections, dust masses inferred from single-temperature models tend to systematically underestimate the true dust masses. This effect is most pronounced for dust with a low mass-weighted temperature yet a broad temperature distribution, in which case $M_\mathrm{d}$ may be underestimated by up to $\sim0.6\,\mathrm{dex}$. However, a systematic underestimate of dust masses in galaxies at high redshift poses further challenges to theoretical models of dust production, which often already struggle to reproduce the massive dust reservoirs inferred from single-temperature models (e.g., \citealt{dayal2022,sommovigo2022b,esmerian2022,esmerian2024,dicesare2023,Choban2024}). 

If early dust is primarily produced through supernovae, as has been argued in several works (e.g., \citealt{todini2001,ferrara2016,dayal2022,esmerian2022}), a systematic underestimate of $M_\mathrm{d}$ implies that high-redshift SNe need to produce even more dust -- by up to a factor of $4\times$ in the most extreme case explored in our models. Observations of local supernova remnants \citep{niculescu-duvaz2022,shahbandeh2023} suggest dust yields up to $\sim 0.7\,M_\odot$ could plausibly be produced, although effective yields could be much lower after destruction from the reverse shock is considered ($<0.1\,M_\odot$; \citealt{bianchi2007,galliano2018}; see also the recent review by \citealt{Schneider23}). If high-redshift SNe therefore have yields similar to the aforementioned local ones, this likely requires invoking a top-heavy initial mass function to explain high-redshift dust masses through SNe alone (c.f., \citealt{michalowski2015,lesniewska2019}). 

Precisely because of the high necessary SN yields -- even when adopting the plausibly underestimated dust masses from single-temperature fitting -- several theoretical works have argued that ISM dust growth is likely needed also at high redshift \citep{popping2017,vijayan2019,palla2024}. Indeed, massive galaxies at $z\gtrsim6$ may already be sufficiently metal-enriched (c.f., \citealt{rowland25,shapley2025}) that ISM growth likely plays a significant role \citep{asano2013,Choban2024}. However, dust growth timescales are hard to constrain observationally, and would likely need to be extremely rapid to explain the high dust masses seen at $z\sim7$ \citep{witstok2023_beta,algera2025}, yet also the apparent lack of dust at even higher redshifts ($z\gtrsim8$; \citealt{ciesla2024,algera2025,burgarella2025,mitsuhashi2025}).

On the theoretical side, dust process modeling on microscopic scales relies on nucleation theory \citep{Dwek1998}, which is now starting to be tested both in the laboratory, and through dedicated numerical simulations of likely cosmic dust analogs (e.g., \citealt{demyk2017,bossion2024}). Theoretical dust production models generally predict and/or assume different characteristic timescales for dust growth within $0.3-30$ Myr (e.g., \citealt{dayal2022}) modulo an additional inverse dependence on metallicity \citep[see the review by][ for a compilation of different dust growth prescriptions provided in the literature]{Schneider23}. Since dust processes -- shattering, growth, accretion -- are dependent on the density and radiation field properties, implementing them into state-of-the-art hydrodynamical simulations capable of modeling the multiphase ISM -- and particularly the cold phase -- is paramount to fully address the impact of such properties on dust-to-gas, dust-to-metal or dust-to-stellar mass ratios. 

Nowadays, several hydrodynamical simulations indeed include a physically motivated evolutionary model for the dust mass \citep{Graziani2020, Li2021, Choban2022,Choban2024}, and in some cases further incorporate the evolution of (a simplified) grain size distribution \citep{McKinnon2018, Aoyama2017, Aoyama2020, Hou2019, Romano2022Dust, Narayanan2023, Dubois2024}. These models highlight the non-linear evolution of the dust-to-gas and dust-to-metal ratios, which vary according to ISM conditions where dust grains grow and are destroyed.
Although their consensus is similar to that of semi-analytical models -- i.e., that current observations are pushing the requirements of dust production from stellar sources, that growth is required, and overall that the measured dust-to-stellar mass ratios are pushing the limits of dust production constraints -- a definitive conclusion remains to be reached. This is largely because the high-resolution simulations capable of modeling the colder ISM phases typically lack a sufficient number of analogs to the massive, UV-bright sources probed by ALMA observations at $z>6$ \citep{Choban2024b,Narayanan2025}.

To summarize the above discussion, the existence, nature, and severity of the so-called `dust budget crisis' remain far from clear. However, a consistent underestimate of galaxy dust masses due to the effects of multi-temperature dust certainly does not help to reconcile observations and models. On the other hand, multi-temperature dust is only one of several uncertainties affecting observational determinations of high-redshift dust masses, as we proceed to discuss in the next section.


\subsection{How uncertain are our dust masses really?}
\label{sec:discussion_Mdust_uncertainty}

Up to this point, we tackled the specific question of how multi-temperature dust distributions affect the observationally inferred dust masses and IR luminosities of high-redshift galaxies. In Section \ref{sec:results}, we found that -- in general -- the true and fitted $M_\mathrm{d}$ are in good agreement, though in extreme cases dust masses might be underestimated by up to $\sim0.6\,\mathrm{dex}$. To place this number into context, it is worthwhile to investigate \textit{other} sources of uncertainty that also plague studies investigating the nature and abundance of dust at high redshift.

We focus here on the best-case scenario where one has `Super-ALMA-like' observations of a $z\sim7$ galaxy of one's choosing. In other words, it is assumed that the MBB dust temperature can be measured to a reasonably high accuracy. If not the case, for instance when only a single-band detection is available, the inferred dust mass will be systematically uncertain by $\sim1\,\mathrm{dex}$ depending on whether a dust temperature of $T_{\rm d,MBB} \approx 30\,\mathrm{K}$ is assumed (c.f., \citealt{algera2024,algera2024b}), versus $T_{\rm d,MBB} \approx 80\,\mathrm{K}$ (c.f., \citealt{bakx2020}).\footnote{Certainly, better $T_{\rm d}$ estimates than taking the extrema of the full range of observed temperatures might be available even in the case of a single-band FIR continuum detection, leveraging e.g., the \cii{}-based method from \citet{sommovigo2021,sommovigo2022}, and/or the energy-balance method from \citet{inoue2020,fudamoto2023}. Nevertheless, we here aim to be conservative and take full inventory of potential uncertainties in $M_{\rm d}$.} Assuming then that $T_{\rm d,MBB}$ is constrained, there are three key -- and very much interrelated -- uncertainties beyond the multi-temperature dust whose effect we already explored: the dust model $\kappa_\nu$, the dust emissivity index $\beta_{\rm d}$, and the opacity $\tau_\nu$. Effectively, this boils down to the fact that at a fixed observing frequency $\nu$ the normalization of a modified blackbody $N_\nu$ is not solely the dust mass, but instead (in the optically thin limit) the factor $N_\nu \propto \kappa_\nu M_{\rm d}$ (Eq.\ \ref{eq:MBB}). Unsurprisingly -- but perhaps worth reiterating -- the inferred dust mass thus depends sensitively on the assumed dust model (Eq.~\ref{opacity}). 

Generally, an assumed dust model prescribes a value of $\beta_{\rm d}$ to be used for internal consistency. Indeed, \citet{bianchi2013} already pointed out that it is typically incorrect to adopt a fixed normalization of the dust model $\kappa_0$ while at the same time varying $\beta_{\rm d}$ when fitting for the dust mass. Nevertheless, this is common practice at high redshift (e.g., \citealt{algera2024,algera2024b} and many others that shall not explicitly be named), and in fact we did exactly this when fitting to the Super-ALMA mock observations in Section \ref{sec:results_superALMA}. While perhaps not completely justified, it is clear that $\beta_{\rm d}$ does vary on a source-by-source basis (e.g., \citealt{witstok2023_beta,algera2024b,liao2024,tripodi2024}), and naturally we would like to capture such variation when fitting. Given that the precise nature of dust at high redshift is not known, and that there is unlikely to be a `one size fits all' dust model anyway, varying $\beta_{\rm d}$ in one's MBB fit therefore appears excusable, as long as the adopted dust model is clearly stated.

What remains, then, is to investigate how much the adopted dust model affects inferred dust masses at high redshift. Remaining agnostic to whichever the `correct' dust model may be, we collect a few values recently adopted in the high-redshift literature, and assess how they impact the inferred $M_{\rm d}$. This is complementary to works such as \citet{fanciullo2020}, who use a physically-motivated dust model obtained from laboratory measurements of cosmic dust analogues, and investigate how this affects the inferred dust masses of distant galaxies. Here, we simply aim to assess the scatter between different high-redshift works, which are important to consider before making direct comparisons between studies.

\begin{itemize}
    \setlength{}
    \item $(\kappa_0, \nu_0) = (10.41\,\mathrm{cm^2\,g}^{-1}, 1900\,\mathrm{GHz})$, based on \citet{draine2003} \citep[adopted in e.g.,][]{bakx2021,sommovigo2022,algera2024,mitsuhashi2024}
    \item $(\kappa_0, \nu_0) = (8.94\,\mathrm{cm^2\,g}^{-1}, 1900\,\mathrm{GHz})$ based on \citet{hirashita2014} \citep[adopted in e.g.,][]{schouws2022,witstok2022,witstok2023_beta} 
    \item $(\kappa_0, \nu_0) = (10.0\,\mathrm{cm^2\,g}^{-1}, 1200\,\mathrm{GHz})$ based on \citet{hildebrand1983} \citep[adopted in e.g.,][]{hashimoto2019}
    \item $(\kappa_0, \nu_0) = (30.0\,\mathrm{cm^2\,g}^{-1}, 2998\,\mathrm{GHz})$ based on \citet{inoue2020} \citep[adopted in e.g.,][]{sugahara2021}
    \item $(\kappa_0, \nu_0) = (0.45\,\mathrm{cm^2\,g}^{-1}, 250\,\mathrm{GHz})$ based on \citet{beelen2006} \citep[adopted in e.g.,][]{tripodi2024}
    \item $(\kappa_0, \nu_0) = (0.77\,\mathrm{cm^2\,g}^{-1}, 353\,\mathrm{GHz})$ based on \citet{dunne2000} \citep[adopted in e.g.,][]{dacunha2021}.   
\end{itemize}

Converting each to $\nu_0 = 1900\,\mathrm{GHz}$ with a fixed $\beta_\mathrm{d} = 2$ yields a range of $\kappa_0 = 8.94 - 26.0\,\mathrm{cm^2\,g}^{-1}$. Given that the inferred dust mass is inversely proportional to the adopted value of $\kappa_0$, the resulting differences in $M_{\rm d}$ span nearly $0.5\,\mathrm{dex}$. Of course, following the above discussion, differences in inferred dust masses can be larger when $\beta_{\rm d}$ is kept free in the fit -- particularly when its fitted value is significantly different from $\beta_{\rm d} = 2$, in which case different dust models may yield dust masses that differ by up to $\sim1\,\mathrm{dex}$. 

Having discussed the effects of the dust model and emissivity index, we briefly turn towards a final source of significant uncertainty, the optical depth $\tau_\nu$. The wavelength where the dust becomes optically thick does not depend on the dust model, since it is proportional to the product of $\kappa_\nu M_{\rm d}$ (that is, $\tau_\nu \propto \kappa_\nu M_{\rm d} / R^2$ for a homogeneous medium with $R$ being the size of the galaxy). Dust masses inferred from optically thin models yield larger values than those assuming optically thick dust (e.g., \citealt{algera2024b}), but without sufficient data at wavelengths bluewards of the peak, the optical depth is difficult to constrain. Based on the brightness of emission lines such as \ciil{} and \oiiil{}, which high-resolution observations show to be co-spatial with the bulk of the dust reservoir (\citealt{rowland2024}; Rowland et al.\ in preparation), \citet{algera2024b} have argued that REBELS-25 -- one of the dustiest galaxies currently known at $z > 7$ -- is unlikely to be globally optically thick beyond $\lambda_\mathrm{thick} \lesssim 65 - 90\,\mu\mathrm{m}$. The difference in recovered dust mass w.r.t. optically thin models is $\lesssim0.2 - 0.4\,\mathrm{dex}$ \citep{algera2024b}, and therefore even smaller for less dusty -- i.e., more typical -- galaxies at high redshift. As such, the effects of optical depth on inferred dust masses have a roughly comparable uncertainty to the effect of multi-temperature dust. \\

To summarize the above discussion, even when the dust SED of a hypothetical $z\sim7$ galaxy is robustly constrained through multi-band far-IR observations, dust masses remain uncertain due to the unknown dust model (by $\sim0.5-1\,\mathrm{dex}$), the effects of multi-temperature dust ($M_{\rm d}$ underestimated by $\lesssim0.3 - 0.5\,\mathrm{dex}$), and the effects of optically thick dust ($M_{\rm d}$ overestimated by $\lesssim0.2 - 0.4\,\mathrm{dex}$, though potentially higher for particularly obscured sources such as SMGs). While this may sound discouraging -- as even with near-perfect knowledge of the shape of the dust SED, one may still not know $M_{\rm d}$ -- it is important to keep in mind that these systematic uncertainties are unlikely to mask any observed trends between $M_{\rm d}$ and other parameters (barring any unfortunate yet unlikely conspiracies). For instance, any observed trends between dust-to-gas ratio and metallicity should be robust as long as a consistent dust model is adopted to convert dust masses inferred from different studies. Such a trend will still be able to reveal the underlying physical and evolutionary mechanisms, even if the true normalization may be uncertain at the $\sim0.5\,\mathrm{dex}$ level. Furthermore, comparing dust yields from local SNe with dust masses in high-redshift galaxies remains fruitful as long as the same dust model is adopted and the aforementioned uncertainties are accounted for. Any mismatch between local and high-$z$ SN yields will then either suggest different dust production efficiencies at early and late times, and/or true differences in the dust properties of local and distant galaxies -- both of which are most certainly interesting conclusions.

\section{Conclusions}
\label{sec:conclusions}
Inferring the global dust properties -- essentially the dust mass, emissivity index and obscured SFR -- of high-redshift galaxies from far-infrared continuum fitting is challenging due to the limited availability of multi-wavelength sub-mm data. As a result, their dust SEDs are often approximated by a single-temperature modified blackbody despite observational and theoretical evidence that this is an oversimplification. In this work, we assess the accuracy of single-temperature models in constraining $M_{\rm d}$ and the infrared luminosity $L_{\rm IR}$ (and thus the obscured SFR) by constructing realistic dust SEDs using a physically motivated analytical model, where the dust temperature probability distribution function (PDF) is described by a simple skewed normal distribution.
We validate this model against a high-resolution, radiative transfer post-processed hydrodynamical simulation from the SERRA suite \citep{pallottini2022}, a simulated quasar host galaxy from \citet{diMascia22}, and a handful of well-studied high-redshift galaxies and quasars at $z\sim6 - 7$ \citep[][Algera \& Herrera-Camus et al.\ in preparation]{tripodi2023,algera2024b}. These comparisons demonstrate that our model successfully reproduces the complexity of mass-weighted and luminosity-weighted temperature distributions (in the simulations), as well as global IR SEDs (in both simulations and observations).

We systematically explore how variations in the mean temperature ($\bar{T}_{\rm d}$), width ($\sigma_T$), and skewness ($\zeta$) of the dust temperature distribution affect the recovery of global dust properties. We thereby sample the constructed dust SEDs for a fiducial $z=7$ galaxy in various ALMA bands, considering both a typical three-band setup (bands 6, 7 and 8), and a more optimistic albeit achievable eight-band setup (all of bands 3 - 10). We fit the resulting dust SEDs using a Monte Carlo Markov Chain-based framework that assumes the commonly adopted single-temperature approximation. Our key findings are:

\begin{itemize}[topsep=2pt, itemsep=3pt, leftmargin=15pt, label=$\square$]
    \item In our three-band ALMA setup, observationally-inferred dust masses underestimate the true values by $\sim0.1 - 0.5\,\mathrm{dex}$ (Fig.\ \ref{fig:basicALMA}). The effect is largest for galaxies with a cold mass-weighted temperature ($\langle T_{\rm d} \rangle_\mathrm{M}\sim30-50\,\mathrm{K}$), but where the temperature PDF is broad. This could correspond to the typical dust properties of metal-rich galaxies such as those found in REBELS \citep{algera2024,algera2024b,rowland25}. Infrared luminosities are generally well-recovered, though also potentially underestimated (by up to $\lesssim0.4\,\mathrm{dex}$) for high mass-weighted temperatures ($\langle T_{\rm d}\rangle_\mathrm{M} \gtrsim 50\,\mathrm{K}$) with broad temperature PDFs. Regardless, the magnitude of any systematic errors is comparable to the observational uncertainties for both $M_{\rm d}$ and $L_\mathrm{IR}$. 

    \item In our optimistic eight-band setup, systematic offsets start to dominate over the measurement uncertainties (Fig.\ \ref{fig:superALMA}). As in the three-band setup, dust masses are underestimated by $\sim0.1 - 0.6\,\mathrm{dex}$, with larger offsets for broader temperature PDFs. IR luminosities, on the other hand, are not strongly systematically underestimated (generally $\lesssim 0.3\,\mathrm{dex}$), in particular for narrow dust temperature PDFs.
    
    \item Observationally inferred dust emissivity indices $\beta_{\rm d}$ are systematically flatter than the intrinsic values due to multi-temperature dust (Fig.\ \ref{fig:beta_hist}). However, the effect is limited ($\Delta \beta_{\rm d} \lesssim 0.2$) when high-S/N observations in bands sampling the Rayleigh-Jeans tail are available. Nevertheless, when steep values of $\beta_{\rm d}$ are observed, this could point towards intrinsic differences in grain properties compared to the local Universe, such as in the grain size distribution.
    
\end{itemize}

Overall, we find that the effects of multi-temperature dust can lead to dust masses being significantly underestimated, while infrared luminosities for typical galaxies are unlikely to be substantially biased. However, when high-S/N dust continuum measurements across a large number of ALMA bands are available -- as will hopefully be the case for large numbers of high-redshift galaxies in the future -- systematics related to multi-temperature dust will begin to exceed even the generally robust measurement uncertainties obtained from Bayesian single-temperature fitting techniques. 

Taking inventory of the various uncertainties in high-redshift dust mass measurements, we find that those intrinsic to uncertainties in the assumed dust model ($\kappa_\nu$) dominate. The adopted model, which depends on the underlying and inherently unknown dust grain properties -- though which we might be able to probe in the future through dust attenuation curve studies \citep{witstok2023_2175,markov2024,fisher2025,Lin25,mckinney2025} -- can lead to systematic differences of $\gtrsim0.5\,\mathrm{dex}$ between dust mass measurements across different studies. For typical $z\sim7$ galaxies, this is larger than the uncertainty from multi-temperature dust ($\lesssim0.5\,\mathrm{dex}$) and optically thick dust ($\lesssim0.3\,\mathrm{dex}$) -- especially considering that these two effects bias dust mass measurements in opposite directions. Care must thus be taken when comparing dust mass measurements across different studies.

In a future work, we aim to incorporate our dust temperature model directly into a Bayesian fitting code, which should yield more robust parameter uncertainties than single-temperature models and yield direct observational estimates of mass- and luminosity-weighted dust temperature PDFs. This will be especially relevant in the era of \textit{PRIMA} \citep{moullet2023}, which is set to bridge the wavelength gap between \textit{JWST}/MIRI and ALMA and probe the mid-IR and far-IR dust emission of large numbers of high-redshift galaxies. Furthermore, we foresee the application of our simple analytical dust model in semi-analytical models of galaxy evolution. In a future study, we will work on physically motivated scaling relations 
to derive the free parameters in our multi-temperature model from global galaxy properties.

\section*{Acknowledgments}
The authors would like to thank the referee for their helpful feedback on the paper. The authors would also like to thank Fabio di Mascia, Lapo Fanciullo, Chris Hayward, Hanae Inami, and Andrea Pallottini for useful discussions. We also thank Fabio di Mascia and Andrea Pallottini for sharing their simulation outputs for comparison.

\section*{Data Availability}
No data were newly generated as part of this article.



\bibliographystyle{mnras}
\bibliography{main}


\bsp	
\label{lastpage}
\end{document}